\documentclass[useAMS,usenatbib,useAMS]{mnras}

\usepackage{newtxtext,newtxmath}

\usepackage[T1]{fontenc}

\DeclareRobustCommand{\VAN}[3]{#2}
\let\VANthebibliography\thebibliography
\def\thebibliography{\DeclareRobustCommand{\VAN}[3]{##3}\VANthebibliography}

\usepackage{graphicx}	
\usepackage{amsmath}	
\usepackage{rotating}
\usepackage{times}
\usepackage{longtable}
\usepackage{color}

\definecolor{darkgreen}{rgb}{0.0,0.5,0.0}
\definecolor{orange}{rgb}{1,0.5,0}
\definecolor{darkred}{rgb}{0.75,0.,0.2}
\definecolor{magenta}{rgb}{0.8,0,0.8}
\definecolor{purple}{rgb}{0.5,0,0.5}
\definecolor{gray}{rgb}{0.5,0.6,0.7}
\definecolor{darkblue}{rgb}{0.1,0.1,0.75}

\newcommand{\msun}{\,{\rm M}_{\sun}}

\title[Internal kinematics of GCs and dwarfs]{The accretion history of the Milky Way. II. Internal kinematics of globular clusters and of dwarf galaxies}

\author[Francois Hammer et al.]{
Francois Hammer$^{1}$\thanks{E-mail:francois.hammer@obspm.fr},
Jianling Wang $^{1,2}$, Gary A. Mamon$^{3}$, Marcel S. Pawlowski$^{4}$, Yanbin Yang$^{1}$,
 \newauthor
Yongjun Jiao$^{1}$, Hefan Li$^{2}$, Piercarlo Bonifacio$^{1}$,
Elisabetta Caffau$^{1}$, 
Haifeng Wang$^{5}$  
\\
$^{1}$GEPI, Observatoire de Paris, Paris Sciences et Lettres, CNRS, Place Jules Janssen 92195, Meudon, France.\\
$^{2}$CAS Key Laboratory of Optical Astronomy, National Astronomical Observatories, Beijing 100101, China\\
$^{3}$Institut d'Astrophysique de Paris (UMR7095: CNRS \& Sorbonne Universit\'e), 98 bis Bd Arago, 75014, Paris, France\\
$^{4}$Leibniz-Institut fuer Astrophysik Potsdam (AIP), An der Sternwarte 16, D-14482 Potsdam Germany\\
$^{5}$CREF, Centro Ricerche Enrico Fermi, Via Panisperna 89A, I-00184, Roma, Italy
}

\date{Accepted 2023 XXX Received 2023 July 13; in original form 2023 July 13}

\pubyear{2023}

\begin{document}
\label{firstpage}
\pagerange{\pageref{firstpage}--\pageref{lastpage}}
\maketitle

\begin{abstract}
We study how structural properties of globular clusters and dwarf galaxies  are linked to their orbits in the Milky Way halo.  From the inner to the outer halo, orbital energy increases and stellar-systems gradually move out of internal equilibrium: in the inner halo, high-surface brightness globular clusters are at pseudo-equilibrium, while further away, low-surface brightness clusters and dwarfs appear more  tidally disturbed. Dwarf galaxies are the latest to arrive into the halo as indicated by their large orbital energies and pericenters, and have no time for more than one orbit. Their (gas-rich) progenitors likely lost their gas during their recent arrival in the Galactic halo. If dwarfs are at equilibrium with their dark matter (DM) content, the DM density should anti-correlate with pericenter.  However, the transformation of DM dominated dwarfs from gas-rich rotation-supported into gas-poor dispersion-supported systems is unlikely accomplished during a single orbit. We suggest instead that the above anti-correlation is brought by the combination of ram-pressure stripping and of Galactic tidal shocks. Recent gas removal leads to an expansion of their stellar content caused by the associated gravity loss, making them sufficiently fragile to be transformed near pericenter passage. Out of equilibrium dwarfs would explain the observed anti-correlation of kinematics-based DM density with pericenter without invoking DM density itself, questioning its previous estimates.
Ram-pressure stripping and tidal shocks may contribute to the dwarf velocity dispersion excess. It predicts the presence of numerous stars in their outskirts and a few young stars in their cores.

\end{abstract}




\section{Introduction}

Milky Way (MW) globular clusters (GCs) and dwarf
galaxies are unique systems, because they are sufficiently close to allow
estimating their intrinsic properties on one hand, and their 3D bulk motions
thanks to the combination of redshifts and proper motions(PMs). Considerable
efforts have been made in estimating their structural \citep[and references
  therein]{Munoz2018}, kinematic \citep[and references therein]{Simon2019},
and orbital \citep{Li2021,Hammer2023} properties. We have recently shown that
GC half-light radii are inversely proportional to their bulk orbital
  energies, and therefore mostly determined by MW tides, which impact depends strongly
on the number of pericenter passages \citep[hereafter Paper~I]{Hammer2023}. \\
  
The main intrinsic properties of GCs and dwarf galaxies are their half-light
radii ($r_{\rm half}$)\footnote{$r_{\rm half}$ is the half-projected-light radius, which is usually called \emph{effective radius}, though we kept the first appellation for consistency with Paper~I.}, 
 their velocity dispersion ($\sigma_{\rm los}$ and for
some GCs, $\rm \sigma_{pm}$ from proper motions), and their total
$V$-band luminosities ($L_{V}$). The
main orbital properties are the total energy, the angular momentum, the
pericenter and the eccentricity of their orbits. Here, we consider the
correlations between structural and orbital properties
of both GCs and dwarfs to investigate how their host affects them. As in
Paper~I, we consider all stellar systems of the MW halo, i.e., GCs and dwarf galaxies. Our motivation is two-fold:
first, these systems have a similar range in stellar mass, and second, as
pointed out by \citet{Marchi-Lasch2019}, there are similarities between
structural properties of ultra faint dwarfs and low surface brightness
GCs. This leads us to distinguish three categories of populations, the
high-surface brightness ({\rm  HSB-GCs}, with $\log(\rm SB/L_{\odot}
pc^{-2}>2$), the low-surface brightness (LSB-GCs, with  $\log(\rm
SB/L_{\odot} pc^{-2}<2$) globular clusters (see Figure 1 of Paper~I, where SB is defined as the mean surface-brightness within the half-light radius), and the dwarf galaxies with lower surface brightness than that of {\rm  LSB-GCs}.  \\

  Paper~I has also established an empirical relation between the
infall lookback time and the logarithm of the orbital energy, for
which the time  has
been calibrated from the age of GCs associated to the different merger events
that occurred in the MW
\citep{Kruijssen2019,Kruijssen2020,Malhan2022}. This empirical
relation is in excellent agreement with theoretical predictions from
cosmological simulations\footnote{However a very different analysis
  \citep{Barmentloo2023} has been recently published, which is discussed in
  Appendix~\ref{sec:Barmentloo}.} \citep{Rocha2012}. It suggests that  most dwarfs
 arrived recently ($<$ 3 Gyr) in the MW halo, because, e.g., their orbital energy is larger than that of Sgr, which is known to have been accreted 5$\pm$ 1 Gyr ago. Such a recent arrival for dwarfs could have a considerable impact on their past evolution and even on their dark matter (DM) content. During such a small elapsed time, dwarf orbiting at large distances and with large pericenters (such as those of classical dwarfs but Sgr) would not make more than one orbit. Because their progenitors far from the MW halo have to be gas-rich dwarfs \citep{Grcevich2009} dominated by rotation, their properties may be governed by their gas losses during their infall into the MW halo. This mechanism has been studied intensively and proved to be effective by \citet{Mayer2006} for DM dominated progenitors, assuming the transformation was done in a Hubble time scale. Here, dwarf progenitors would have much less time to be transformed into dispersion supported dwarfs, so their DM content should be more limited than what was assumed by \citet{Mayer2006}.\\

Section~\ref{sec:data} describes the orbital and intrinsic properties of samples of GCs and of dwarf galaxies, and Section~\ref{sec:MWmass} discusses how dwarf eccentricities and infall times depends on the MW total mass. Section~\ref{sec:corr} compares the intrinsic GC properties, and how they correlate\footnote{Throughout the manuscript we have used a Spearman's rank correlation $\rho$ that does not assume any shape for the relationship between variables; the significance and associated probability of $\rho$ have been tested using  $t$= $\rho$ $\sqrt{(n-2)/(1-\rho^2)}$, which is distributed approximately as Student's $t$ distribution with $n-$2 degrees of freedom under the null hypothesis.} with orbital properties, showing that HSB-GCs are in pseudo-equilibrium with MW tides, while LSB-GCs appear to be much less in equilibrium. For dwarf galaxies, Section~\ref{sec:dwarfs} presents a strong correlation between a simple combination of intrinsic parameters ($\sigma_{\rm los}^2/r_{\rm half}$) and the pericenter, which is at the origin of many correlations shown throughout the Paper. Section~\ref{sec:discussion} shows that these correlations can be explained if recently infalling dwarfs have been stripped of their gas and tidally shocked by the MW. It also provides a theoretical calculation for the effect, which is confirmed by numerical simulations (Sect.~\ref{sec:disc_simu}). It leads to important predictions, e.g., of a tiny young stellar component in their cores (Sect.~\ref{sec:disc_SFH}), of the presence of a stellar halo surrounding most MW dwarfs (Sect.~\ref{sec:disc_outskirts}), and on some limitations about the DM content of MW dwarfs (Sect.~\ref{sec:disc_DM}). Section \ref{sec:conclusions} summarizes the results and conclusions of this study.

\section{Orbital and intrinsic properties of GCs and dwarfs}
\label{sec:data}

\subsection{Globular clusters}

In the following, we consider the data for 156 {\rm  GCs} from  \citet{Baumgardt2017,Baumgardt2018,Baumgardt2020,Baumgardt2021,Sollima2017}, which include their intrinsic parameters such as half-light radii and velocity dispersions. Proper motions from Gaia EDR3 are taken from \citet{Vasiliev2021}. Tables A1 and A2 in Appendix of \citet{Hammer2023} provide the resulting orbital parameters (velocities and orbital radii) and their error bars for a MW model following \citet[see also a more detailed description in \citealt{Jiao2021}]{Eilers2019}. 

\subsection{Dwarf galaxies}

 For dwarf galaxies, we are using the data from Gaia EDR3 \citep[see their Tables for each MW mass models]{Li2021} using the same prescriptions as for GCs for deriving orbital parameters. Appendix~\ref{sec:parameters} describes the adopted values for dwarf intrinsic parameters, and their references (see Table~\ref{tab:struct}). 
\subsection{Sample definition}
Since the goal of this series of Papers is to examine the impact of the MW on its halo inhabitants, we have not considered GCs that are associated to the LMC. The same applies for dwarf galaxies and we have excluded Carina II, Carina III, Phoenix II, Horologium I, Hydrus I, and Reticulum II, which are associated to the LMC on the basis of their relative proper motions \citep{Erkal2020,Patel2020}. This reduces the potential impact of a massive LMC, which has not been considered in the following. This leaves us with 26 dwarf galaxies having measurements of their internal kinematics ($\sigma_{\rm los}$), to which we may add Sgr, the status of which is quite special given its large system of tidal tails \citep{Ibata2001}. In the following we will only consider the sample of 26 MW dwarfs, without Sgr (see Table~\ref{tab:struct} of Appendix~\ref{sec:parameters}).

\subsection{Milky Way mass models and how they affect GC and dwarf orbits}
\label{sec:MWmass}
\citet{Eilers2019} modeled the MW rotation curve using a mass model that
includes a bulge, a thick and a thin disk following \citet{Pouliasis2017}
corresponding to a total baryonic mass of 0.89 $\times$ $10^{11}M_{\sun}$,
and a halo dark matter component. Here, the halo is represented by four
different models. The first one is a NFW model \citep*{Navarro1996} that
requires a cut-off radius fixed at $R_{\rm vir}$= 189 {\rm kpc} to avoid
infinite mass (see details in Paper~I), and for consistency with orbital calculations tabulated in \citet{Li2021} and in Paper~I. This model is very similar than the fiducial {\it MWPotential14} from \citet{Bovy2015} which has been widely used.
We are also considering the whole range of MW mass models able to fit the
rotation curve (see, e.g., \citealt{Jiao2021}), by considering three different
Einasto \citep[see Table~\ref{tab:mass} and also \citealt{Retana-Montenegro2012}]{Einasto1965}
profiles, which include the largest and smallest MW mass (model high and low
mass, HM and LM, respectively) and a median mass (MM, see
Table~\ref{tab:mass}). Einasto$_{\rm HM}$ model shares a similar total mass than the \citet{McMillan2017} model, while conversely to the later, it is consistent with the MW rotation curve from \citet{Eilers2019}. Einasto$_{\rm MM}$ model is coming from the analysis of MW rotation curves and globular clusters made by \citet{Wang2022}. Einasto$_{\rm LM}$ is rather similar (but with a slightly higher total mass) to that of \citet{Ou2023} who generated the first Einasto modeling based on the MW rotation curve obtained from Gaia EDR3.

\begin{table}
 \caption{Properties of the four Milky Way dark-matter mass models to be associated to the baryonic disks and bulge (see text).}
\label{tab:mass}
 \begin{tabular}{lcccc}
\hline
  Quantity (units) & Einasto$_{\rm HM}$ & NFW & Einasto$_{\rm MM}$ &
  Einasto$_{\rm LM}$\\
\hline
$M_\mathrm{DM}\ (10^{11}\msun)$ & 14.1 & 7.2 &  4.2 & 1.9\\
$M_\mathrm{tot}\ (10^{11}\msun)$ & 15 & 8.1 &  5.1 & 2.8\\
$r_\mathrm{200}\ (\mathrm{kpc})$ & 236 & 189 & 164 & 135\\
$r_{-2}$ (kpc) & 29.89  & 14.80 & 12.81 & 9.73 \\
Einasto index & 6.33 & -- & 3.00 & 1.67 \\
$\chi_\nu^2$ & 1.57 & 1.27 & 1.21 & 0.72 \\
\hline
 \end{tabular}
 \parbox{\hsize} {$M_{\rm DM}$ and $M_{\rm tot}$ are defined within the virial
     radius $r_\mathrm{200}$.  $r_{-2}$ is the radius where the logarithmic slope of the
   density profile is equal to $-2$ for direct comparison between the NFW and
   Einasto profiles.
    The last row ($\chi_\nu^2$) provides their goodnesses in fitting the MW rotation curve (see calculation details in \citealt{Jiao2021})
}
\end{table}

For each MW mass model, GC and dwarf eccentricities are calculated from {\sc galpy} \citep{Bovy2015} using Monte Carlo realizations. For elliptical orbits the eccentricity is:
\begin{equation}
  {\rm ecc}=\frac{R_{\rm apo}-R_{\rm peri}}{R_{\rm apo}+R_{\rm peri}} \ .
  \label{ecc_ell}
  \end{equation}
  
  For hyperbolic orbits, we estimated the eccentricity after deriving the angle 2$\theta$ between asymptotes:
 \begin{equation}
  {\rm ecc}=-\frac{1}{\cos(\theta)} \ .
  \label{ecc_hyp}
  \end{equation}

 \begin{figure}
\includegraphics[width=3.5in]{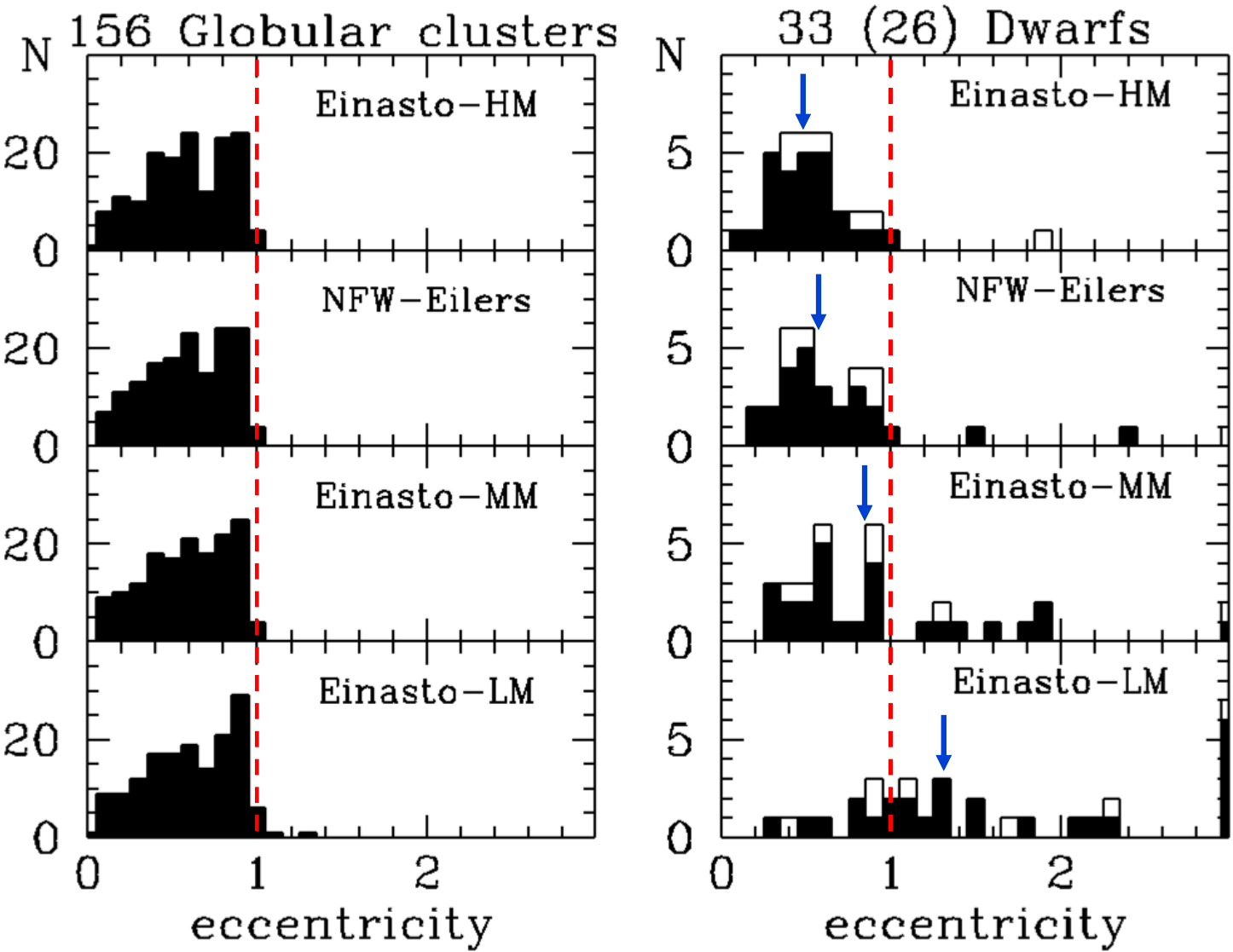}
\caption{Histograms of the eccentricity distribution of GCs (\emph{left}) and dwarf
  galaxies (\emph{right}) for four MW mass models (decreasing total mass from top to
  bottom) that are consistent with the MW rotation curve
  \protect\citep{Jiao2021}. \emph{Open} and  \emph{filled histograms} represent all the 33 dwarfs,
  and a subsample of 26 of them having their internal velocity dispersion
  measured. The \emph{blue arrows} in the right panels indicate the median
  eccentricities of 0.478, 0.581, 0.829, and 1.301 from the top to the
  bottom. The \emph{vertical red-dashed line} delimits ecc=1. 
} 
\label{fig:ecc}
\end{figure}
 
Figure~\ref{fig:ecc} shows how the orbital eccentricity depends on the MW
mass models. The distribution of eccentricities of 156 GCs is very stable
when decreasing the MW mass from 15 to $5.1 \times 10^{11}\msun$, i.e., all of them follow elliptical and bound orbits. It is only by adopting the smallest MW mass ($2.8\times10^{11}\msun$) that  the Pyxis and Terzan 8 orbits becomes hyperbolic (ecc=1.3 and 1.07, respectively), while the orbits of three other GCs (Pal 3, Eridanus, and Arp 2) have eccentricities just below 1.\\

This contrasts with the net increase of dwarf eccentricities when the MW mass
decreases, from the top to the bottom right panels of
Figure~\ref{fig:ecc}. Dwarf galaxies show a smaller median eccentricity than that of GCs for high MW mass models, while two-thirds of them are on hyperbolic orbits for the lowest MW mass model (Einasto$_{\rm LM}$).   Table~\ref{tab:ecc} of Appendix~\ref{sec:parameters} gives dwarf eccentricities for each of the four adopted MW mass models of Table~\ref{tab:mass}.
\\

Figure~\ref{fig:ecc} illustrates that by using GCs to characterize the MW mass, one would find values close or larger than that of the Einasto-MM model (e.g., see \citealt{Wang2022} and references therein).  If considering dwarfs as MW satellites, one would automatically derive large masses for the MW, which suggests that MW mass determinations are strongly affected by the choice of adopted priors. It implies that the total dynamical mass of the MW derived from the Gaia DR2 rotation curve \citep{Eilers2019,Jiao2021} is still an unknown within a large range of values. \\

Another illustration of the impact of the MW mass choice is given by the
determinations of the infall time for MW dwarfs. According to
\citet{Boylan-Kolchin2013}, satellites with the most recent infall have the largest
orbital energy (or the smallest binding energy), which is well illustrated in
Figure 1 of \citet{Rocha2012}. This follows the expectations of the
\emph{onion skin} model of \cite{Gott1975}. It has prompted many studies
\citep{Rocha2012,Fillingham2019,Miyoshi2020,Barmentloo2023} to use dedicated
zoomed simulations to directly compare MW dwarfs with simulated
satellites. If considering large MW masses, dwarf galaxies would have smaller
energy than objects near the escape velocity lines, which unavoidably leads
to large infall lookback times, as found by the above studies that considered
$M_\mathrm{tot}/ (10^{11}\msun)$= 19,  10-25 (average 17 from the ELVIS
suite, \citealt{Garrison-Kimmel2014}), 15.4, and 10-20, respectively. Right panels of
Figure~\ref{fig:ecc} show that by decreasing the MW mass below this range, more and more dwarfs become unbound, leading to small infall lookback times. It suggests that the inferred infall time is thus dependent on what total MW mass has been assumed. 
\\

This is why, in Paper~I and in this paper, we have adopted a different approach by estimating only relative infall times, after comparing MW dwarf orbital energies to robust estimates of the infall times for past merger events in the MW \citep{Kruijssen2020,Malhan2022}, i.e., independently of the MW mass.  Our constraints on the MW dwarf infall time are coming from a comparison with assumed Gaia-Sausage-Enceladus (8-10 Gyr ago), and Sgr (4-6 Gyr ago) infall times. Since the orbital energy of MW dwarfs are larger than that of the latter events, their infall epochs are expected to be more recent (see Figure 6 of Paper~I).

In the following we will try to identify which relationship between intrinsic and orbital properties is independent on the MW mass model, either for dwarfs or for GCs. 


\section{Globular cluster properties}
\label{sec:corr}

\begin{figure}
\includegraphics[width=3.5in]{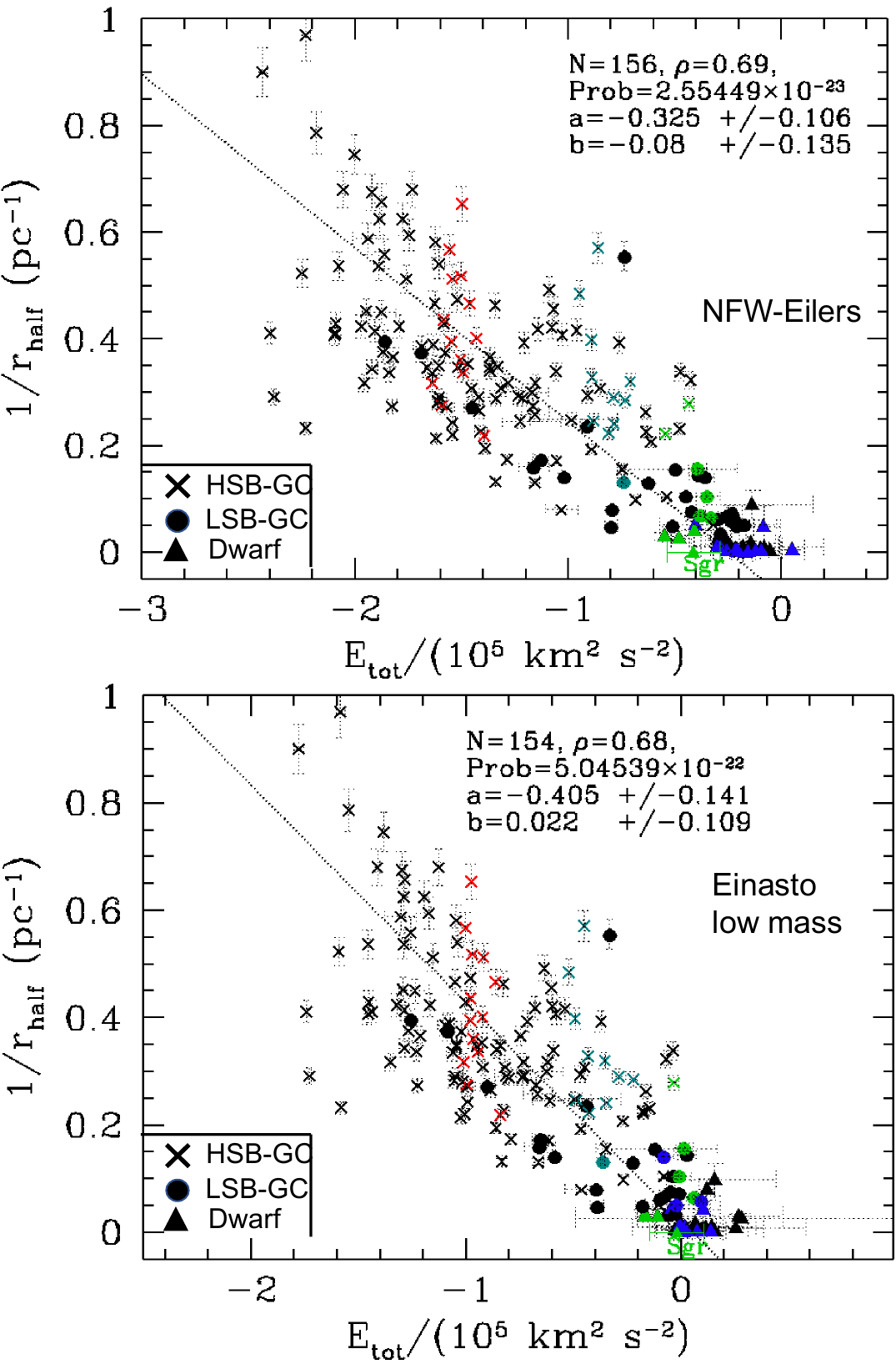}
\caption{Inverse of the half-light radius versus total orbital energy, for
  {\rm  HSB-GCs} (\emph{crosses}), {\rm  LSB-GCs} (\emph{full dots}), and dwarfs (\emph{triangles}). 
 Identified structures by \citealt{Malhan2022} and \citealt{Kruijssen2020} are represented by different color codes (Kraken: red; Gaia-Sausage-Enceladus: cyan; Sgr: green). Dwarf galaxies and LSB-GCs belonging to the VPOS \citep{Pawlowski2012} are shown with blue colors.  The \emph{top} (\emph{bottom}) panel shows 156 (154) GCs and 33 dwarfs after
 calculating the total energy from the NFW (Einasto-low mass) model of the
 MW. The location of Sgr is indicated by a \emph{green triangle}.  The
 \emph{dotted line} shows the fit of the GC points, and the corresponding
 slope (a) and rest (b), as well as the significance of the Spearman rank correlation are given on the top-right of each panel. One sigma error bars on orbital quantities such as pericenter and total energy are based on Monte Carlo calculations assuming Gaussian distributions for PM and radial velocity errors.} 
\label{fig:E-rh}
\end{figure}

\subsection{The correlation between half-light radius, pericenter radius and total energy}

In Paper~I, we have shown that the half-light radius scales as the inverse of
the total orbital energy ($E_{\rm tot}$) for the NFW model of the MW \citep[see also
  Table~\ref{tab:mass}]{Eilers2019,Jiao2021}. Figure~\ref{fig:E-rh}
illustrates that this anti-correlation trend is highly  significant (probability
of a coincidence smaller than $10^{-20}$). We have tested the slope of the logarithmic relation between $r_{\rm half}$ and $E_{\rm tot}$ for the 4 MW mass models of Table~\ref{tab:mass}, which ranges from --1.1$\pm$0.3 to --1.18$\pm$0.3. We then adopt\footnote{Slopes (a) provided in the top-right side of each panel of Figure~\ref{fig:E-rh} are coming from the linear fit of $E_{\rm tot}$ with 1/$r_{\rm half}$, and are then different that those from the logarithmic relation.} $-1$ for simplicity, i.e., $E_{\rm tot}$ $\propto$ 1/$r_{\rm half}$ in Figure~\ref{fig:E-rh}. 
It shows that the GC size depends on the number of previous passages at pericenter, and that smaller stellar systems are associated to early infall into the MW  (see Paper~I). This relation does not depend on the MW model as it is illustrated by comparing top and bottom panels of Figure~\ref{fig:E-rh}, for which GCs share very similar locations. Figure~\ref{fig:E-rh} also illustrates how identified structures (see colored points) have almost a single energy value, suggesting the relation between their formation epoch and the energy described in Paper~I.

\begin{figure}
\includegraphics[width=3.5in]{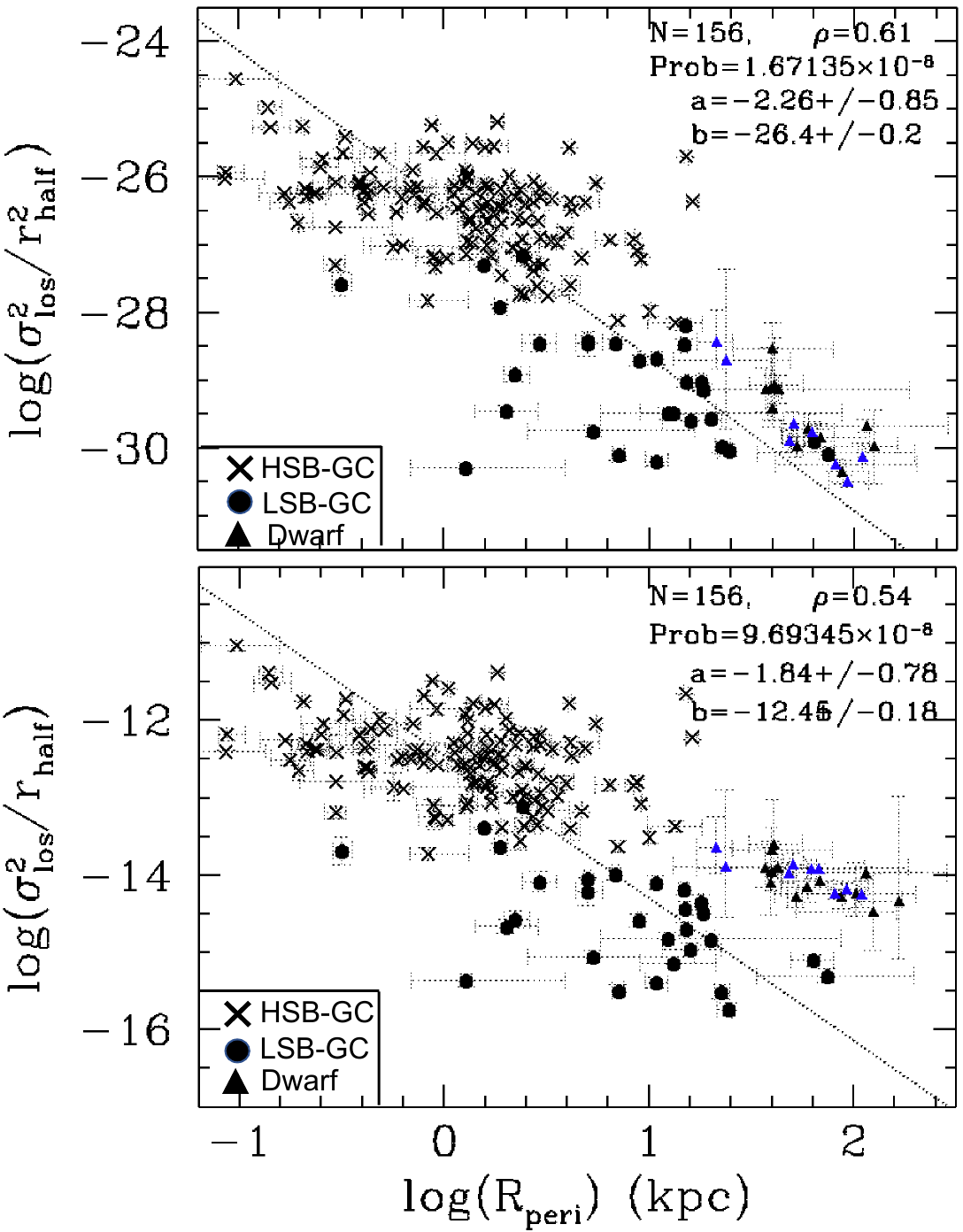}
\caption{$\sigma_{\rm los}^2 r_{\rm half}^2$ (proportional to a 3D density) and $\sigma_{\rm los}^2 r_{\rm half}$ (proportional to a surface density) of GCs versus $R_{\rm peri}$. Same symbols than for Figure~\ref{fig:E-rh}, except that here only VPOS dwarfs have kept their (blue) color. One sigma error bars on orbital quantities such as pericenter and total energy are based on Monte Carlo calculations assuming Gaussian distributions for PM and radial velocity errors.} 
\label{fig:s2srh_Rp_GCs}
\end{figure}

For GCs we adopt the same relation as \citet{Baumgardt2017} between the
dynamical mass 
(or total mass inside $r_{\rm half}$) and the velocity dispersion and the
half-light radius, which has been established by \citet{Wolf2010}, under the
assumption of self equilibrium and constant line of sight velocity
  dispersion:
 \begin{equation}
  \frac{M_{\rm tot}(r_{\rm half})}{1\msun}= 930 \,
  \left(\frac{\sigma_{\rm los}}{1\, {\rm km\, s}^{-1}}\right)^2 \, \left(\frac{r_{\rm
      half}}{1\rm  pc}\right) \ ,
  \label{Wolf}
  \end{equation}
where $\sigma_{\rm los}$ has been estimated within $r_{\rm half}$.
 
By dividing the total mass by $r_{\rm half}^2$ and by $r_{\rm half}^3$, one
obtains quantities proportional to the averaged surface density
$\Sigma_{\rm tot} \propto \sigma_{\rm los}^2/r_{\rm half}$
   and to the 3D density ($\rho_{\rm tot} \propto \sigma_{\rm los}^2/r_{\rm half}^2$) inside $r_{\rm half}$, respectively. Figure~\ref{fig:s2srh_Rp_GCs} shows how the two latter quantities depend on  pericenter\footnote{Here we choose to use pericenter instead of the total orbital energy since we need to establish logarithmic relations to identify their scaling power; one may recall that many dwarfs and even few GCs can be unbound and with positive energy, contrary to most inhabitants of the MW halo (see Figure~\ref{fig:ecc}). }. The resulting anti-correlations (see straight-dotted lines) are not unexpected for GCs. We have shown that their pericenter (and angular momentum) is well correlated with their energy (see Paper~I), while Figure~\ref{fig:E-rh} shows how the energy correlates with the half-light radius. The latter correlation is likely responsible for the strong, but quite scattered relation between surface-density and 3D-density with pericenter.\\ 

\begin{figure}
\includegraphics[width=3.5in]{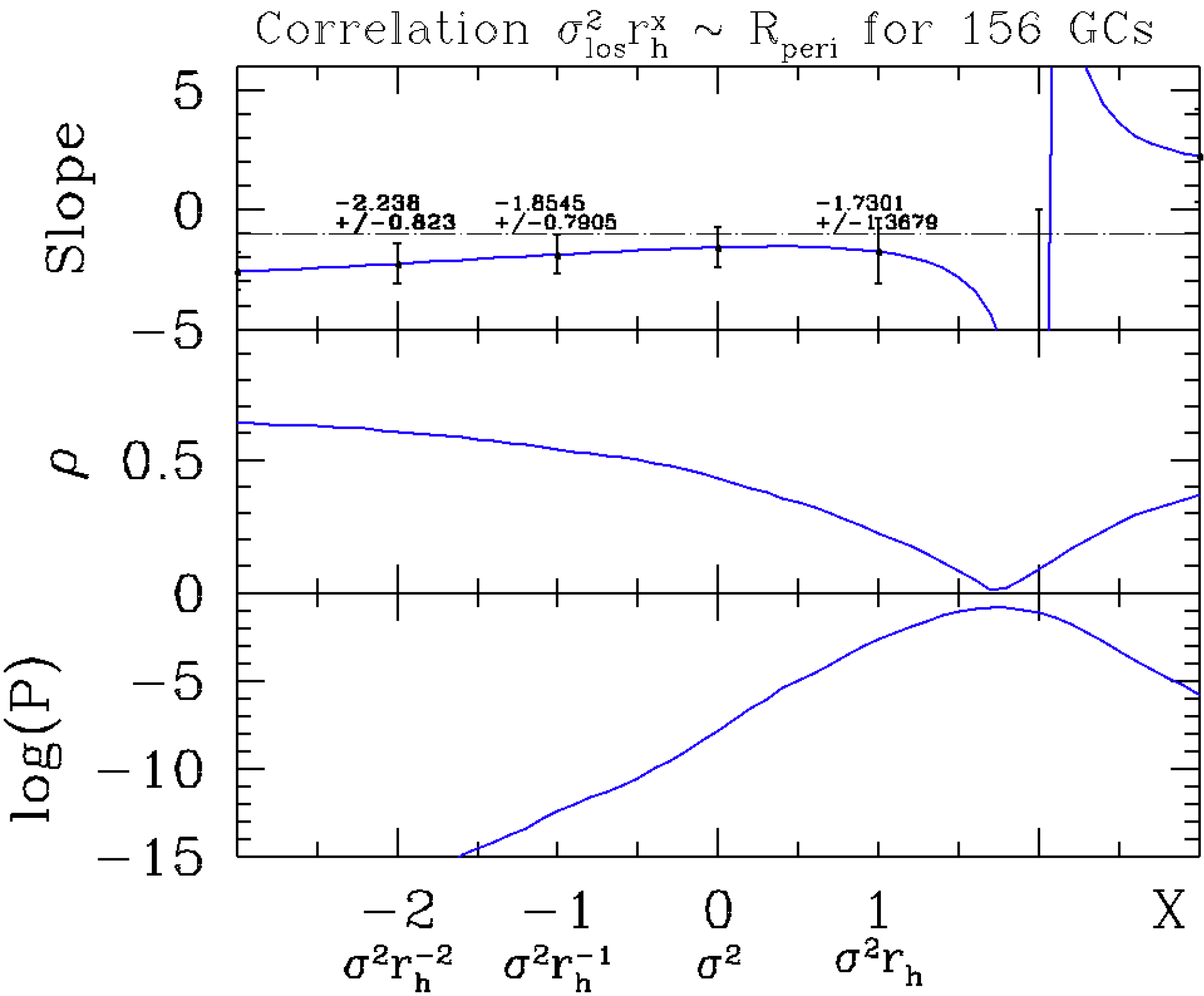}
\caption{Slope, correlation $\rho$ (its absolute value), and associated probability (in
  logarithmic scale) of the correlation between $\sigma_{\rm los}^2 r_{\rm
    half}^X$ and $R_{\rm peri}$  for $X$ (in abscissa) taking values from $-3$
  to +3. Each quantity is shown on the $X$ axis. The NFW model has been assumed for calculating $R_{\rm peri}$, but curves for other models cannot be distinguished from it.
 }
\label{fig:sig2rhX_Rp_GCs}
\end{figure}

Figure~\ref{fig:sig2rhX_Rp_GCs} shows how slope, correlation significance,
and probability are changing with X when considering the relation between
$\sigma_{\rm los}^2\,r_{\rm half}^X$ and $R_{\rm peri}$ for 156 GCs. For $X=
1$, the former represents the total mass that barely correlates with $R_{\rm
  peri}$, while the anti-correlation coefficient ($\rho$) increases for
 increasing $|X|$. This is expected because
the pericenter should be directly related to the orbital energy (see Paper~I), which is
proportional to the half-mass radius (see Figure~\ref{fig:E-rh}), so the velocity dispersion should be a
more minor factor.

\subsection{How tidal shocks affect an orbiting stellar system}
Tidal shocks exerted on GCs near their pericenter have been theoretically
described for orbits passing close to the bulge or through the disk
\citep{Aguilar1988,Gnedin1999}. \citet{Aguilar1988} showed that the average energy increase of a star after integrating over the whole GC has been calculated to be:
\begin{equation}
\Delta E=\frac{4}{3}\, \eta \left({G M_{\rm MW}(R_{\rm peri}) \over R_{\rm peri}^2}\right)^2\,
\left({r_{\rm rms}\over V_{\rm peri}}\right)^2 \, ,  
 \label{Eq:Aguilar}
\end{equation}
where $V_{\rm peri}$ is the velocity at pericenter, $\eta$ is a dimensionless parameter that accounts for our lack of understanding of the details of the tide (deviations from the impulsive approximation\footnote{ $\eta$ is always equal or less than 1, except in the outermost layers
      of the satellite due to resonances. It absorbs the effects of adiabatic
      invariants, and of an extended perturber (the impulse approximation involves
      a point-like perturber). Both effects shrink the magnitude of the effect of the
      tidal shock.}), and:

\begin{equation}
 r_{\rm rms}^2 = 3 \, r_{\rm half, 3D}^2,
 \label{Eq:Keenan}
\end{equation}
with:
\begin{equation}
r_{\rm half, 3D}=  \frac{4}{3} r_{\rm half},
 \label{Eq:Wolfe}
\end{equation}
the latter comes from \citet{Wolf2010} after transforming the theoretical
$r_{\rm half, 3D}$ into the observed $r_{\rm half}$ calculated from the
stellar surface density.
The problem with the rms radius is that it may diverge for 
  models whose outer density slope is $>-5$ (including the Plummer model, often used to represent
  the density profiles of GCs). Only the \citet{King1966} model and truncated
  models would have finite rms radius. 
  One possibility  is to limit the calculation of the rms radius to the bound particles.

One extreme case is tidal disruption, where, to first order, the energy
  impulse per unit mass from the MW tide matches the binding energy per unit
  mass $3/2 \sigma^2$, where $\sigma$ is the average one-dimensional velocity
  dispersion of the system (but see \citealt{vandenBosch+18} for a more
  detailed analysis that shows the resilience of stellar systems to tides). Then,
one can combine Eqs.~\ref{Eq:Aguilar}, ~\ref{Eq:Keenan}, and ~\ref{Eq:Wolfe}, yielding:
\begin{equation}
\label{Eq:tidalshocks}
\frac{\sigma^2}{r_{\rm half}^2} = {128\over 27 } \, \eta \left({g_{\rm MW}(R_{\rm peri}) \over V_{\rm peri}}\right)^2  ,   
\end{equation}
where $g_{\rm MW}(R_{\rm peri})$= $G M_{\rm MW}(R_{\rm peri})/R_{\rm peri}^2$ is the gravitational acceleration
exerted by the assumed spherical MW at $R_{\rm peri}$. 
Values of $\rm g_{\rm  MW}$ and of $V_{\rm peri}$ have been derived from Monte Carlo realizations assuming Gaussian distributions for PM and radial velocity errors. These calculations are
based on {\sc galpy} \citep{Bovy2015}, to warrant a correct propagation of errors as well as to consider effects related to axisymmetric disks. 
However, Eq.~\ref{Eq:tidalshocks} corresponds to the specific and extreme case when the energy impulse per unit mass equals the binding energy per unit mass. In the following, we generalize it by considering cases for which the energy impulse only amounts to a fraction $f$ of the binding energy, which applies to many systems that can be affected by tides, but not fully destroyed by them. To simplify, $f$ also accounts for $\eta$, and the latter can shrink below 1 (see Fig. 3 of \citealt{Aguilar1988}). This leads to:

\begin{equation}
\label{Eq:tidalshocks2}
 \frac{\sigma^2}{r_{\rm half}^2} \times f\, = {128\over 27 } \, \left({g_{\rm MW}(R_{\rm peri}) \over V_{\rm peri}}\right)^2  .   
\end{equation}

\begin{figure}
\includegraphics[width=3.5in]{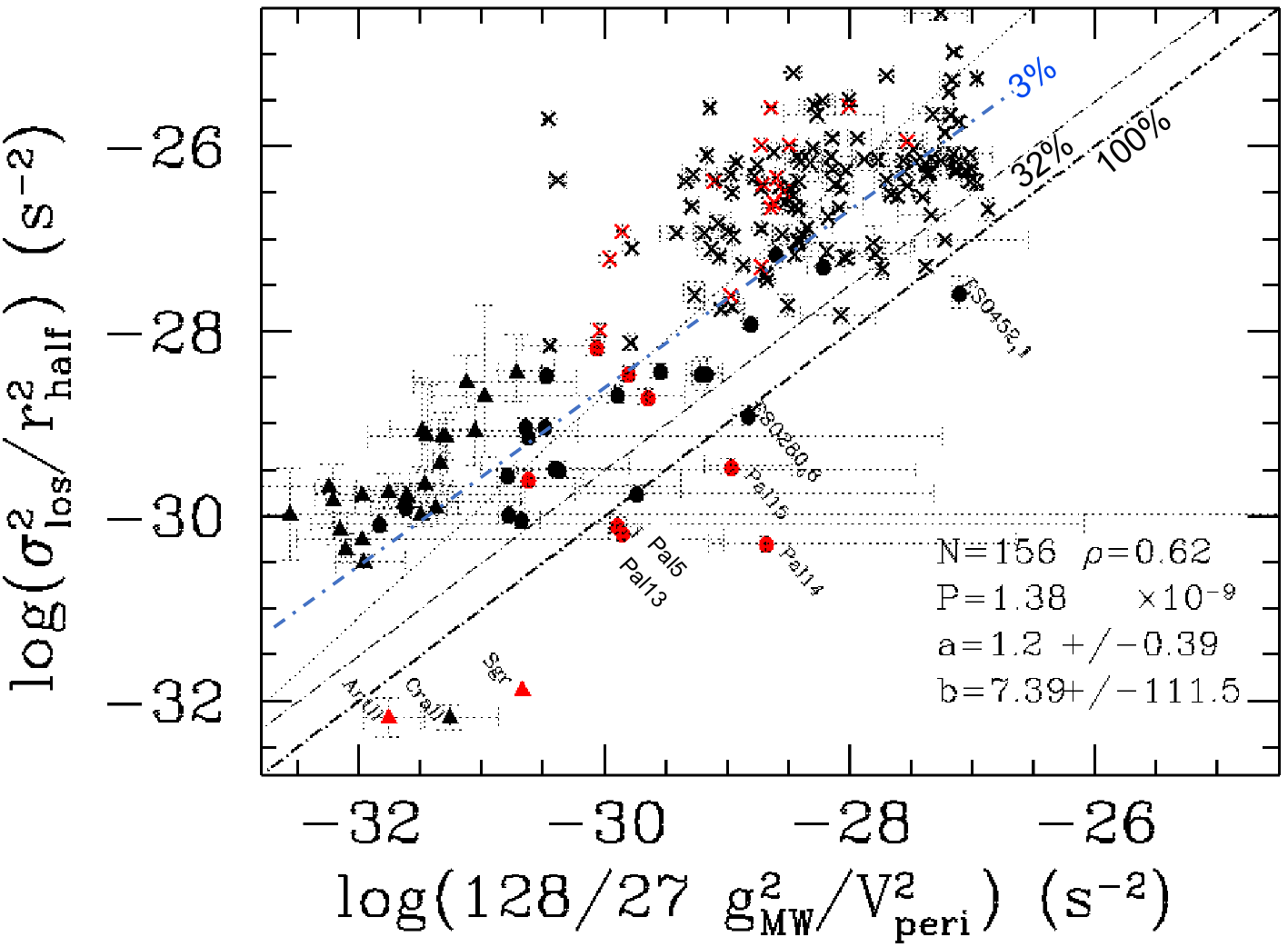}
\caption{Correlation between internal GC properties ($ \sigma_{\rm los}^2 r_{\rm half}^2$,
proportional to 3D density) and external properties  from tidal shock
  theory for a GC \citep{Aguilar1988,Gnedin1999} passing near
  pericenter to be disrupted (see Eq.~\ref{Eq:tidalshocks2}). Same symbols as in
  Figure~\ref{fig:E-rh}, but now indicating in red color the GCs associated
  to a tidal tail \citep[see their Table 3]{Zhang2022}. \emph{Dot-dashed
    lines} correspond to the equality ,
  $f=1$\ (\emph{thick}), to $f=0.32$\
    (\emph{thin}) and to $f=0.032$ (\emph{blue}), respectively, the latter being almost
  confused with the \emph{dotted line} that represents the correlation for
  GCs. All quantities in the abscissa have been extracted at $R_{\rm
  peri}$.
}
\label{fig:tidalshocks}
\end{figure}

Eq.~\ref{Eq:tidalshocks2} compares structural (left) with orbital
(right) quantities, which is illustrated in Figure~\ref{fig:tidalshocks}, in
which we have assumed $\sigma=\sigma_{\rm los}$ in spherical symmetry, a condition that is verified for most GCs.
 Eq.~\ref{Eq:tidalshocks2} accounts for bulge and disk shocks, which are both described by the same formulae, though with different timescales. \\

The left hand of Eq.~\ref{Eq:tidalshocks2} corresponds to the square of the inverse crossing time
\begin{equation}
t_{\rm cross}= {r_{\rm half} \over \sigma}
\end{equation}
for a star inside a stellar system, while the right hand estimates the  inverse-square of a time ($t_{\rm perturber}$) linked to the perturber passage\footnote{The tidal shock time is $t_{\rm shock}$=$R_{\rm   peri} / V_{\rm peri}$, which is equal to $t_{\rm perturber}$ $\times$ $(GM_{\rm MW}/R_{\rm peri}$)/(0.459$V_{\rm peri}^2$), i.e., their ratio is that coming from both sides of the virial theorem, 0.5$V_{\rm peri}^2$=$GM_{\rm MW}/R_{\rm peri}$. If the system is fully virialized the ratio should be very close to 1, though even when not at equilibrium it could not be very different than 1.}, i.e., the MW. 
In the case of a fast perturber, $t_{\rm shocks} \le t_{\rm cross}$ and the impulse approximation may apply
  \citep{Aguilar1985,Gnedin1997}, 
     implying that stellar systems found below the equality
  line in Figure~\ref{fig:tidalshocks} are likely tidally shocked, disrupted,
  or stripped. 
   In such a case, the energy brought by tidal shocks becomes
  equal or larger than the kinetic energy necessary to balance the
  self-gravity of  the stellar system, which becomes dominated by tides.

One would expect that such stellar systems may show tidal tails. Red symbols in Figure~\ref{fig:tidalshocks} indicate GCs for which such tail systems have been identified \citep[see their Table 3]{Ibata2021,Zhang2022}, although these surveys cannot be considered as complete since not all GCs have been scrutinized for tidal tail search. 

\subsection{The different properties of HSB and of LSB-GCs} 

Figure~\ref{fig:tidalshocks} reveals large differences when comparing HSB with LSB-GCs. HSB-GCs appear much more robust against tides, since none (among 127) are below the equality line $f$= 100\%. It contrasts with the 7 (24 \%) LSB-GCs that lie well below the line. Most of the latter possess a known system of tidal tails (4 among 7), while tails are less frequent for LSB-GCs above the line (4 among 22). This furthermore contrasts with HSB-GCs, for which all systems possessing tidal tails are well above the equality line, and even above the $f$= 0.32 line.\\



Figure~\ref{fig:tidalshocks} shows that quantities at both sides of
Eq.~\ref{Eq:tidalshocks2} correlate well (156 GCs, $\rho= 0.62$, $P= 1.4 \times 10^{-9}$). Here we show that it is due to the fact that both quantities anti-correlate with the pericenter. According to the top panel of Figure~\ref{fig:s2srh_Rp_GCs}, $\sigma_{\rm los}^2/r_{\rm half}^2$ evolve as $R_{\rm peri}^{-2.26}$. Most HSB-GCs have their orbits within 15 kpc, for which $M_{\rm MW}$ $\propto$ R (see, e.g., \citealt{Jiao2021}), and then $g_{\rm MW}$=$GM_{\rm MW}/R^2$ $\propto$ $R^{-1}$, and we also find (see Appendix~\ref{sec:Vperi}) that $V_{\rm peri}$ is almost independent to $R_{\rm peri}$. It lets the second hand of Eq.~\ref{Eq:tidalshocks2} following to $R_{\rm peri}^{-2}$. In other words, stellar systems have decreasing pericenter, orbital energy, and half light radii from the left to the right of Figure~\ref{fig:tidalshocks}.  \\ 

HSB-GCS density increases by $\sim$ 3\% at each pericenter passage \citep[see their Fig. 9]{Martinez-Medina2022}, a phenomenon which is called star evaporation\footnote{Stellar systems passing near their pericenters are likely affected by MW tidal shocks, which increase the internal energy of their stars, resulting in the least bound to be expelled. Then, after {\rm  GCs} have lost mass, they contract adiabatically when leaving the pericenter towards the apocenter.} \citep{Binney2008}. Lying on the top-right of Figure~\ref{fig:tidalshocks}, they are experiencing much more pericenter passages than other stellar systems. In Figure~\ref{fig:tidalshocks} their median location is very close to expectations for $f$= 0.03 (see the blue dot-dashed line), which means that they are stellar systems in pseudo equilibrium with the MW potential and tides. This explains why only few of them possess tails. Because they orbit at low radii, many different processes could generate these tails, such as gravitational interactions with other substructures, e.g., giant molecular clouds, spiral arms, and the bar \citep[and references therein]{Ibata2021}.\\

This contrasts with LSB-GCs that appear more fragile due to their much lower densities (300 times on average, see Figure~\ref{fig:tidalshocks}), while at significantly larger pericenters, i.e., their orbits may extend to regions far from the MW disk. 24\% of them are fully tidally shocked, and they are more affected by tides than HSB-GCs, since their locations in Figure~\ref{fig:tidalshocks} are generally well below the $f$= 0.03 line. Conversely to HSB-GCs, their passages to pericenter are relatively rare, and then their properties are not fully shaped by the MW potential, yet. \\
 
HSB-GCs are in pseudo equilibrium with the MW potential. Going further to the bottom-left of Figure~\ref{fig:tidalshocks}, one finds LSB-GCs that are much less at equilibrium. In the next section we examine whether or not dwarfs galaxies that lie at the very bottom-left of Figure~\ref{fig:tidalshocks} are in equilibrium. \\

We have verified that all the above properties, including the relative locations of both GCs and dwarf galaxies relatively to the equality line in Figure~\ref{fig:tidalshocks}, do not change with the MW potential.

\section{Dwarf galaxy properties}
\label{sec:dwarfs}
\subsection{Comparison of dwarf galaxy and LSB-GC properties}
Figure~\ref{fig:tidalshocks} shows that LSB-GCs have more similarities in the investigated properties with dwarf galaxies (full triangles) than with HSB-GCs. First, both populations show a fraction of fully tidally disrupted systems, i.e., those below the equality line in Figure~\ref{fig:tidalshocks}. The three dwarf galaxies below the equality line are Sgr, Crater II and Antlia II. The first is well-known for its gigantic system of tidal streams that surround the MW \citep{Ibata2001}, and the other two, by their extremely low stellar density and large sizes. Antlia II is likely associated with tidal tails, which is also suspected for Crater II\footnote{Both Antlia II and Crater II have been reproduced by a model in which low density systems have lost their gas and are completely out of equilibrium because they are dominated by tides (see \citealt{Wang2023}).} \citep[hereafter Paper~III]{Ji2021,Wang2023}. Second, two LSB-GCs (Pal 3 and Crater) lie in the sequence delineated by dwarfs in Figure~\ref{fig:tidalshocks}. This corroborates \citet{Marchi-Lasch2019} conclusions that structural properties of ultra faint dwarfs and of low surface brightness GCs could be rather similar. \\

However, none of the 24 remaining dwarfs that lie above the equality line show tidal tails\footnote{Carina has been suggested to be with tides \citep{Battaglia2012}, although contamination by LMC debris may discard it \citep{McMonigal2014}, and the 24 dwarf sample does not include Tucana III, which is a unique system by its extremely low pericenter (few kpc) and that possess a tail. Unfortunately there is no robust measurement of Tucana III kinematics.} \citep{Hammer2020}, while some LSB-GCs do.  In addition, most dwarf galaxies do not show a spherical morphology, and their surface brightness are generally fainter than that of LSB-GCs.



\subsection{A strong anti-correlation between structural and orbital dwarf properties}

In Figure~\ref{fig:tidalshocks}, 24 of the 26 dwarf galaxies appear at first glance to be
unaffected by MW tidal shocks due to their location well above the equality
line and the absence of tidal tails. However, one may wonder why they
delineate a sequence that is precisely parallel and offset by $\sim$ +1.5 dex
to the equality line, with a significance of $\rho= 0.66$ ($P = 8.6 \times
10^{-4}$). The same applies after examining the two panels of Figure~\ref{fig:s2srh_Rp_GCs} where dwarf densities correlates with $R_{\rm peri}$ as well as GCs, though being offset from them. To understand this, we propose to examine the core sample of 24 dwarf galaxies obtained after removing Antlia II and Crater II. 
 In the following, we investigate which combination of their intrinsic properties (e.g., $r_{\rm half}$, $\sigma_{\rm los}$, and $L_{V}$) correlates the best with orbital properties (e.g., their pericenter, $R_{\rm peri}$). \\

\begin{figure}
  \centering
  \includegraphics[width=\hsize]{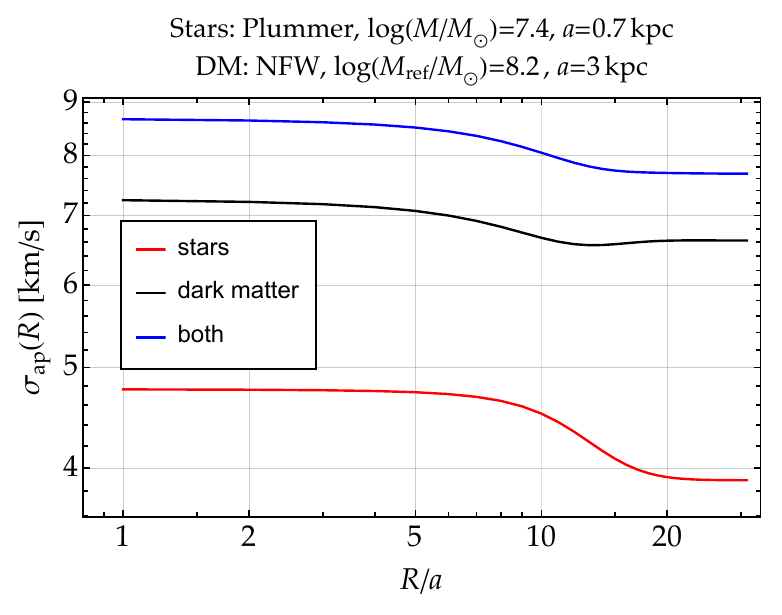}
  \caption{
Contribution of stars and dark matter to the velocity dispersion measured in
a cylindrical aperture, i.e. the average los velocity dispersion, for a
 dwarf such as Fornax.  Here,
$M_{\rm ref} = M_{\rm DM}(a_{\rm dark})$, and isotropic velocities are assumed.
  }
\label{fig:sigap}
\end{figure}

Since tidal forces
scale with distance to the system center, and that dark matter is more
extended than stars, one expects that dark matter can be used to measure the
strength of the tides within dwarf galaxies. We propose to use the dark matter contribution to the
stellar velocity dispersion as the test for tidal theory, and vice-versa. Its  squared value  is:
\begin{equation}
  \sigma_{\rm  DM}^2 \approx \sigma_{\rm los}^2 - \sigma_{\rm stars}^2
  \ , 
\end{equation}
where $\sigma_{\rm stars}$ is the contribution of stars to the stellar
  velocity dispersion (the formula 
  is an approximation because of the neglect of the gas component, which
  appears justified for MW dwarf galaxies).  We measure the contribution of the
  stars to the velocity dispersion, $\sigma_{\rm stars} \equiv \sigma_{\rm ap}(R)$, in a cylindrical
  aperture of radius $R$ by integrating over the cylinder the stellar
  contribution to the radial  velocity
  dispersion of the stars, itself found by integrating the Jeans equation of
  local dynamical equilibrium assuming further isotropic kinematics. This quantity involves a triple integral (one for
  solving the Jeans equation for the radial component of the 3D velocity
  dispersion, one for integrating along the line of sight, and one for
  integrating over the different lines of sight within the aperture). 
We use the single integral exact expression of equation (B7) of
\cite{Mamon&Lokas05}, corrected in \cite{Mamon&Lokas06}, valid for systems
with isotropic velocities
\begin{flalign} 
\sigma_{\rm stars}^2 & \equiv   \sigma_{\rm ap}^2(R) \nonumber \\
  &= {4\pi\,G \over 3 \Sigma(R)}\,
  \left[ \int_0^\infty  r \nu (r)
   M(r)\,{\rm d}r \right. \nonumber \\
   & \qquad \qquad \quad \left. - \int_R^\infty {(r^2-R^2)^{3/2} \over r^2} \nu (r) \,
   M(r)\,{\rm d}r \right] \ ,
  \label{sigmaapiso}
\end{flalign}
where $\nu(r)$ and $\Sigma(R)$ are the 3D and surface stellar number density
profiles, respectively, while $M(r)$ is the total mass profile\footnote{The original triple integral ensures that
  Eq.~(\ref{sigmaapiso}) is a very good approximation to 
  the aperture velocity dispersion for systems with anisotropic velocities.}.
 Figure~\ref{fig:sigap} gives an example for Fornax, assumed to be embedded in a DM halo.

%
\begin{figure}
\includegraphics[width=3.5in]{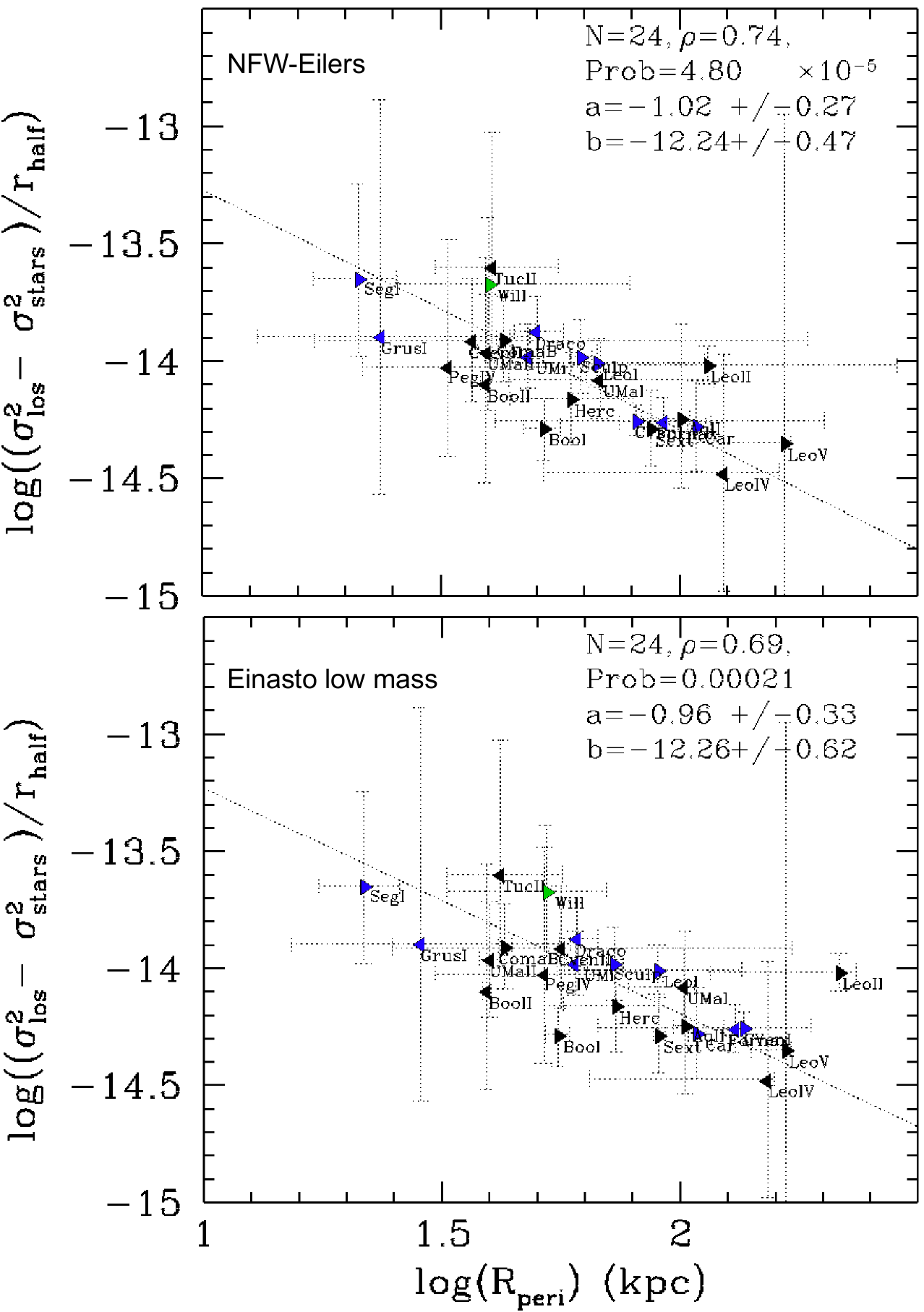}
\caption{ Free-fall tidal shocks (or DM) acceleration (in km $\rm s^{-2}$)
  based on dSph kinematics (($\sigma_{\rm los}^2 - \sigma_{\rm stars}^2) \,
  r_{\rm half}^{-1}$) versus pericenter radius. Data ($\sigma_{\rm
    los}$, $L_{V}$, $r_{\rm half}$) come from Table~\ref{tab:struct} of Appendix~\ref{sec:parameters}.
} 
\label{fig:ffs_Rper}
\end{figure}

The top panel of Figure~\ref{fig:ffs_Rper} shows that when removing
quadratically $\sigma_{\rm stars}$ the correlation reaches a slightly higher
significance, i.e., $\rho= 0.74$ between ($\sigma_{\rm los}^2 - \sigma_{\rm
  stars}^2)/r_{\rm half}$ and $R_{\rm peri}$, which is associated to a low
probability that it occurs by chance, $P= 4.8 \times 10^{-5}$.
\\


\begin{figure}
\includegraphics[width=3.5in]{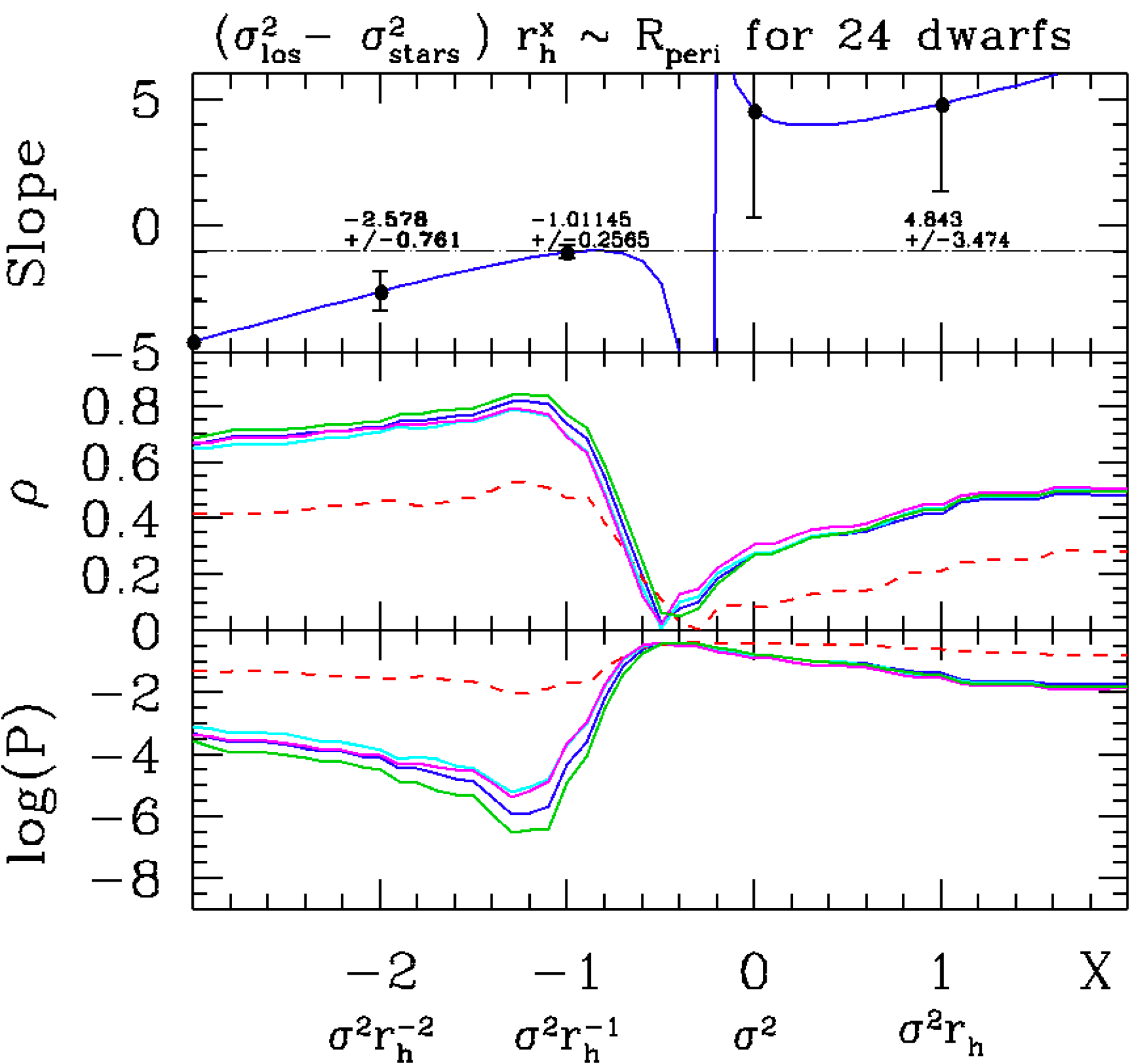}
\caption{
Slope, correlation $\rho$ (its absolute value), and associated probability (in logarithmic scale) of the correlation between $(\sigma_{\rm los}^2 - \sigma_{\rm stars}^2)$ $r_{\rm half}^X$ and $R_{\rm peri}$ (red-dashed line: with  $R_{\rm GC}$) from $X=-3$ to $X=2$. Each quantity is shown on the $X$ axis, and $\rho$, and associated probability are given for the 4 MW mass models (see colored lines).} 
\label{fig:sig2stars_rhX_Rp}
\end{figure}
 
 Figure~\ref{fig:sig2stars_rhX_Rp} shows how slope, correlation significance, and probability are changing with X when considering the relation in logarithmic scale between ($\sigma_{\rm los}^2 - \sigma_{\rm stars}^2)\,r_{\rm half}^X$ and $R_{\rm peri}$ for 24 dwarf galaxies. It shows that the correlation peaked at $X= -1$, which likely drives all correlations at $X < -1$. 
 \\
  
Anti-correlations shown in Figures~\ref{fig:s2srh_Rp_GCs},
~\ref{fig:ffs_Rper}, and ~\ref{fig:sig2stars_rhX_Rp} between  by dwarf galaxy
structural and orbital parameters have already been identified by
\citet{Hammer2019}, who also considered $\sigma_{\rm los}^2-\sigma_{\rm
    stars}^2$, and by \citet{Kaplinghat2019}, who found robust
  anti-correlations between 3D density within 150 pc and pericenter (see
  their fig.~2).  However, \citet{Kaplinghat2019} found that the
anti-correlation vanishes for ultra-faint dwarfs (UFDs). This difference can
be explained because, in contrast to us, they include both Antlia II and
Crater II in the UFD sample, and both objects dominate the relation because
of their extremely low densities (see their Fig. 3). We re-assess that Antlia
II and Crater II (as well as Sgr) are experiencing strong tidal stripping
conversely to the rest of the 24 dwarfs considered here, as it is shown in
Figure~\ref{fig:tidalshocks}. We confirm the \citeauthor{Kaplinghat2019}
result, i.e., by inserting Antlia II and Crater II in
Figures~\ref{fig:s2srh_Rp_GCs}, ~\ref{fig:ffs_Rper}, and
~\ref{fig:sig2stars_rhX_Rp} is sufficient to wash out the anti-correlation,
because they have densities several dex lower than those of other dwarfs.\\

If MW dwarf galaxies were at self-equilibrium with their own gravity, both
$(\sigma_{\rm los}^2 - \sigma_{\rm stars}^2)/r_{\rm half}$ and $(\sigma_{\rm
  los}^2 - \sigma_{\rm stars}^2)/r_{\rm half}^{2}$ would correspond to the
surface and 3D mass densities of the sole dark matter (DM)
component\footnote{This is roughly true for isotropic, isothermal
    systems.}. Looking at the slope of the correlations (see top panel of Figure~\ref{fig:sig2stars_rhX_Rp}), it implies that both
quantities vary as $R_{\rm peri}^{-1}$ and $R_{\rm peri}^{-2.6}$,
respectively. If dwarf galaxies were long-term satellites of the MW, this
could be interpreted as being caused by a 'survivor bias', i.e., satellites
with small pericenter would be those having been shielded against the tidal
forces (see \citealt{Vitral&Boldrini22}), which means that they have to be denser \citep{Kaplinghat2019}, even favoring cusped density profiles in their center \citep{Errani2023}. Assuming that MW dwarfs are long-term satellites, \citet{Robles2021} found that subhalos with small pericenters are indeed denser after a Hubble time evolution, while \citet{Kravtsov2023} did not. \\
 
However, due to their high orbital energy and angular momenta, most dwarf galaxies are stellar systems that arrived late in the MW halo (less than 3 Gyr ago, see Paper~I). It results that they have no time to make one or few orbits in the MW halo, in sharp contrast with a long-term satellite hypothesis.  It thus appears quite enigmatic why their structural parameters such as $r_{\rm half}$ and $\sigma_{\rm los}$ show a correlation with orbital parameters in Figure~\ref{fig:ffs_Rper} and with the tidal-shock characteristic time in Figure~\ref{fig:tidalshocks}.




\section{Discussion}
\label{sec:discussion}

\subsection{Dwarf galaxies with escaping stars affected by MW tides}
\label{sec:escapingstars}
Figure~\ref{fig:sig2stars_rhX_Rp} also shows that the correlations with
pericenter ($R_{\rm peri}$) are much stronger than with galactocentric radius
($R_{\rm GC}$, red line). This suggests that intrinsic properties of dwarf
galaxies are linked to MW tides (see also Sect.~\ref{sec:comb} for a more detailed discussion). This calls for a mechanism related to the expected
progenitor properties of MW dwarfs, a few Gyr ago. Sect.~\ref{sec:disc_DM} discusses the impact of a dwarf recent infall onto their DM properties.\\

Beyond 300 kpc, all dwarfs (but a few, e.g., Cetus and Tucana) are gas-rich,
while they are gas-poor within 300 kpc (except the massive LMC/SMC;
\citealt{Grcevich2009}). If the former are progenitors of the latter, gas-rich dwarfs are expected to be stripped during their infall due to the
ram pressure caused by the Galactic halo gas \citep{Mayer2006}.  The role of
the removed gas during the process could be essential, if it induces a lack of
gravity implying that many stars have to expand \citep{Grishin+21} following a spherical geometry (for an isotropic distribution of initial velocities of stars). The fraction of stars that are lost depends on the total initial mass, on the mass fraction initially represented by the gas, and also on the relative size of the different components in the progenitor. \\

Simulations performed by \citet{Yang2014} have shown that when the gas is
lost near the pericenter passage, it increases the dwarf kinematics, because
many stars are escaping the system and are then affected by the MW
gravity. \citet{Hammer2020} proposed an ideal case for which a fraction
 ($f_{\rm ff}$, ff standing for free-fall) of stars are not affected by the dwarf self-gravity,
but by the MW gravity. Assuming a Plummer profile for the dwarf, they
calculated that the dwarf squared velocity dispersion is increased by:
 \begin{equation}
\Delta \sigma^2 = 2\sqrt{2} \, g_{\rm MW} \, r_{\rm half}  \, f_{\rm ff},
\label{Eq:ffs}
 \end{equation}
where $g_{\rm MW}$ stands for the MW gravity\footnote{This comes from an integration of the velocity dispersion along the line-of-sight for stars assumed to be free falling into the Milky Way gravitational field. However this is quite a simplification since the stars, even the unbound ones that have just escaped the cluster or dwarf, move in the combined gravity of its parent system and the MW.
The first term 2$\sqrt{2}$ would be changed into 2.08$\sqrt{2}$ for a perfect sphere model
  \citep{Hammer2018a}.}. Equation~\ref{Eq:ffs} is over-simplified since in
reality, stars lying in the center may move sufficiently fast that they are
adiabatically invariant to the actions of the moving MW gravitational field
\citep{Weinberg1994,Binney2008}. Some of our simulations (Paper~III) show a
residual core that appears stable against star expansion and MW gravity. This is an adiabatic effect, since the most bound stars have very short
      orbital periods compared to the timescale of the MW flyby. So, for the
      encounter, they appear smeared along their internal cluster orbits, and the
      external perturbation can only kick the orbit as a whole (i.e. mainly linear
      momentum exchange). For the cluster envelope stars, the flyby is impulsive,
      they only cover a fraction of its orbits during the flyby, so the external perturbation
      can do work deforming their orbits (i.e. mainly energy exchange), heating up
      their dynamics and contributing to their expansion and eventually becoming
      unbound.\\

 The consequence would be to change $f_{\rm ff}$ in Equation~\ref{Eq:ffs}
 into a smaller value if stars within $r_{\rm half}$ are less
affected due to adiabatic invariance. Simulations also show that the gas
removal is a turbulent process that affects both gas and star motions during
the time they are bound together, and also affect the velocity dispersion of
stars after gas removal. In the following, we consider this to be accounted
for by the $f_{\rm ff}$ factor, which depends also on the structural
properties of the dwarf. We derive:\\
 \begin{equation}
{\sigma_{\rm los}^2 - \sigma_{\rm stars}^2 \over  r_{\rm half}} = 2\sqrt{2} \, g_{\rm MW} \, f_{\rm ff},
\label{Eq:ffs_gMW}
 \end{equation}

\begin{figure}
\includegraphics[width=3.5in]{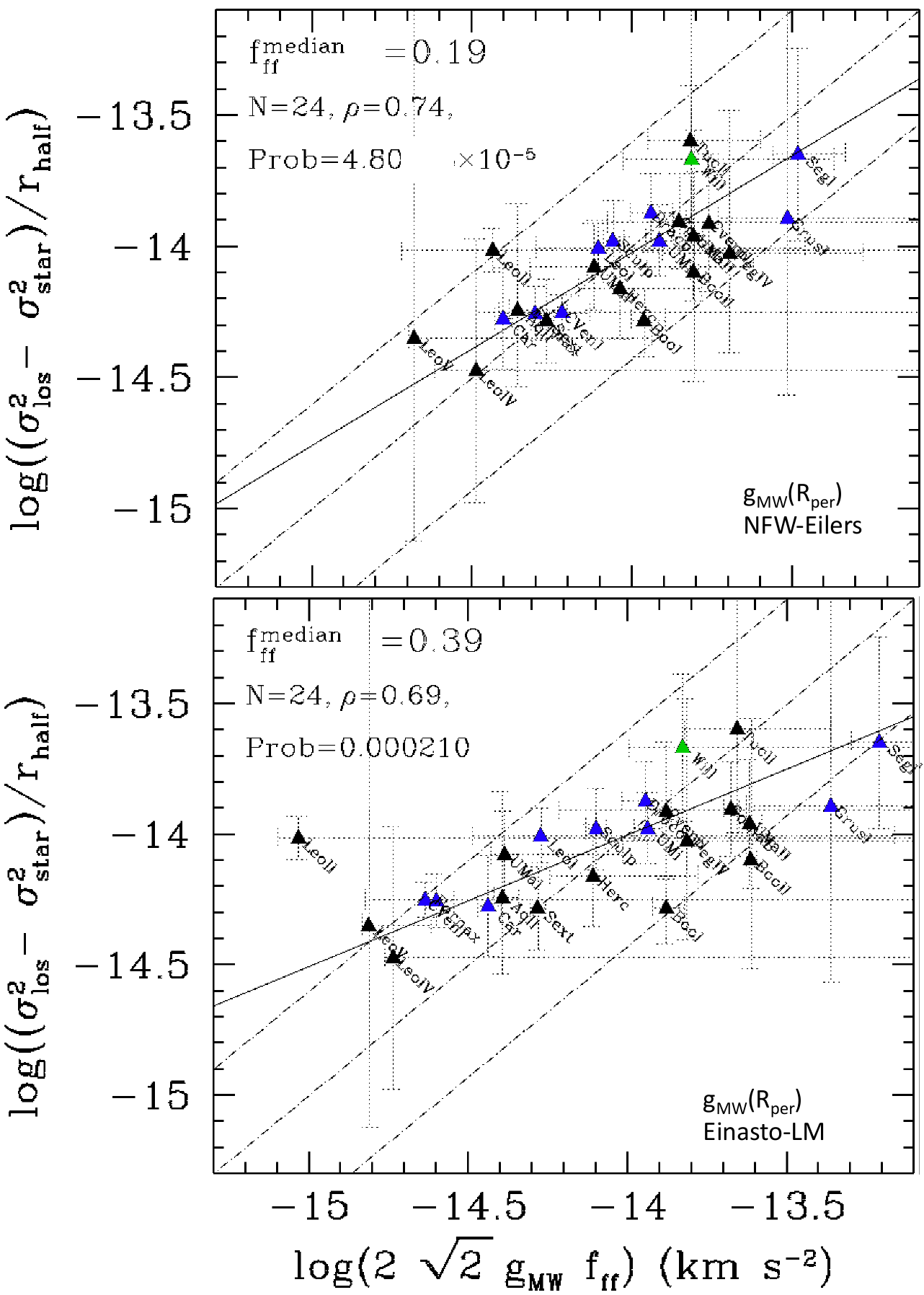}
\caption{ Free-fall tidal shocks (or DM) acceleration (in km $s^{-2}$) based
  on dSph kinematics (($\sigma_{\rm los}^2 - \sigma_{\rm stars}^2) / r_{\rm
    half}$) versus MW gravity at pericenter, for the NFW (\emph{top panel})
  and Einasto low mass (\emph{bottom panel}) MW models. The \emph{solid line}
  represents the best fit of the correlation, whose characteristic numbers
  are provided in the top-left. The line corresponding to the median of
  $f_{\rm ff}$ is the central dot-dashed line, while the two other
  \emph{dot-dashed lines} correspond to the median value, multiplied
  (\emph{top line}) or divided (\emph{bottom line}) by 2.5, respectively.
   } 
\label{fig:ffs_gMW}
\end{figure}

Figure~\ref{fig:ffs_gMW} shows that ($\sigma_{\rm los}^2 - \sigma_{\rm stars}^2)$/$r_{\rm half}$ correlates well with the MW gravity, with a similar correlation strength than that with pericenter. For the NFW MW mass model (top panel) tidal shocks exerted on expanding stars can reproduce the observations if the fraction ($f_{\rm ff}$) of the latter ranges from 0.08 (bottom dot-dashed line) to 0.48 (top dot-dashed line), with a median at 0.19. Classical dwarfs, but Leo II, are mostly near the median value, as well as most dwarfs of the VPOS (blue triangles, see, e.g., \citealt{Pawlowski2012}), but Grus I. \\

When adopting the low MW mass model (bottom panel of Figure~\ref{fig:ffs_gMW}), $f_{\rm ff}$ ranges from 0.16 to 0.98, with a median value of 0.39. However, Leo II appears as an outlier ($f_{\rm ff}$ $>$ 1), which could be caused by the quite large uncertainty on the MW acceleration at pericenter, or alternatively, because it does not obey  Eq.~\ref{Eq:ffs_gMW}.\\

 \begin{figure}
\includegraphics[width=3.5in]{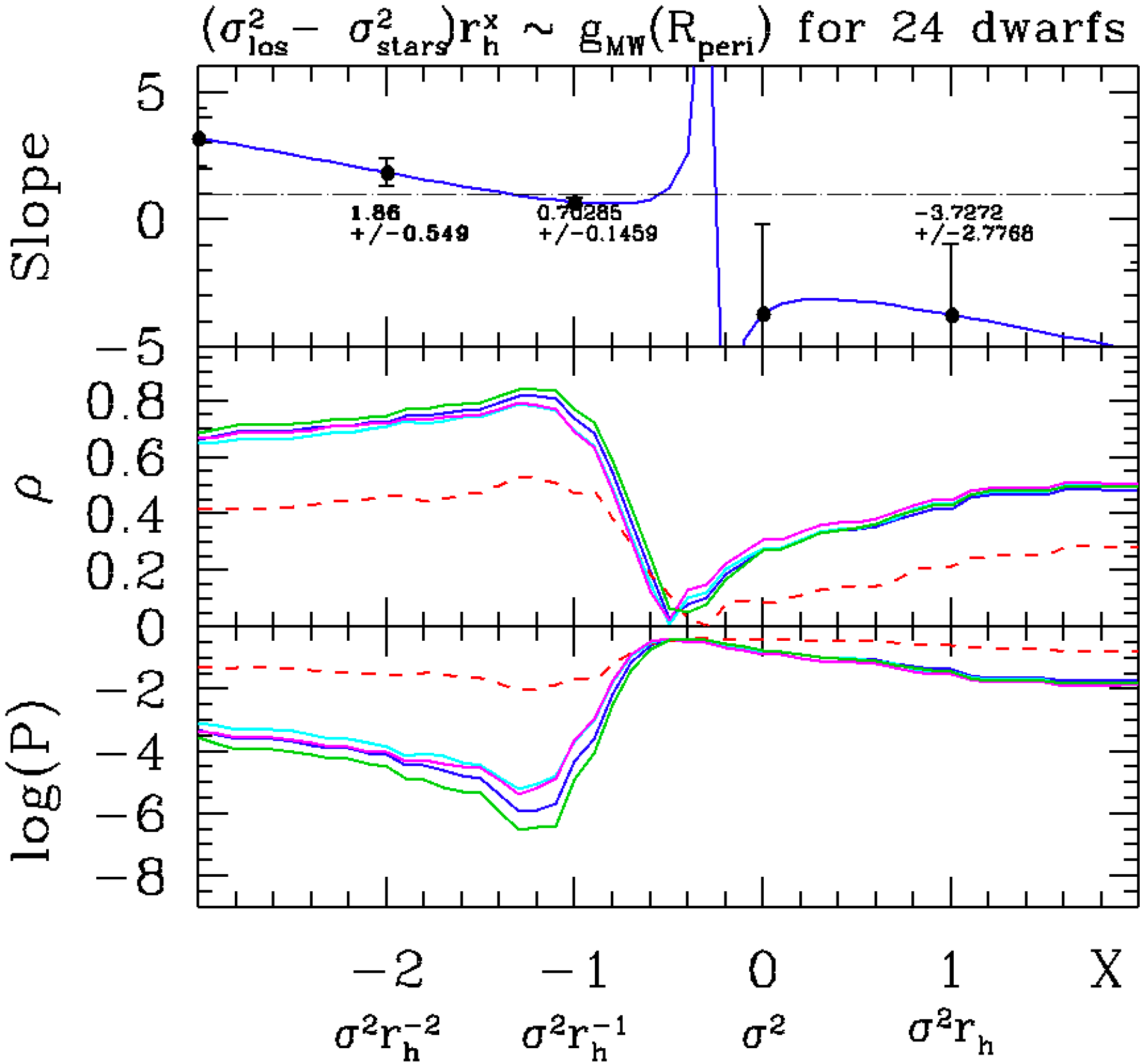}
\caption{Slope, correlation $\rho$, and associated probability (in logarithmic scale) of the correlation between $(\sigma_{\rm los}^2 - \sigma_{\rm stars}^2)$ $r_{\rm half}^X$ and $\rm g_{MW}$ at $R_{\rm peri}$ (red-dashed line: at $R_{\rm GC}$) from $X=-3$ to $X=2$, respectively. Each quantity is shown on the $X$ axis, and associated probability are given for the 4 MW mass models (see colored lines).} 
\label{fig:sig2stars_rhX_gMW}
\end{figure}

Figure~\ref{fig:sig2stars_rhX_gMW} confirms our finding (see
Figure~\ref{fig:sig2stars_rhX_Rp}) that the correlation with $g_{\rm MW}$ is
driven by $X= -1$, which supports the validity of Eq.~\ref{Eq:ffs_gMW}, and then the fact that intrinsic properties such as $\sigma_{\rm los}$ and $r_{\rm half}$ are changing through a temporal sequence, gas removal, star expansion, and then MW tidal shocks. It also shows that the correlation is much improved when adopting MW gravity at pericenter (see the red-dashed line representing gravity at $R_{\rm GC}$), suggesting further a tidal origin for the observed correlations.\\

We notice that Figure~\ref{fig:sig2stars_rhX_gMW} is essentially similar than Figure~\ref{fig:sig2stars_rhX_Rp} probably because in the range of pericenter values, $g_{\rm MW}$ values are simply set by $R_{\rm peri}$ values. However, it provides a good test for the validity of Eq.~\ref{Eq:ffs_gMW}.

\subsection{Comparison with numerical simulations}
\label{sec:disc_simu}
A physical interpretation of the correlation shown in Figure~\ref{fig:ffs_gMW} may need two conditions to be fulfilled at the time dwarfs are observed:
\begin{enumerate}
\item Due to the gas removal and gravity loss a significant fraction of stars are expanding to the dwarf outskirts and are gradually less affected by the dwarf gravity;
\item Gas loss happened at a time quite close from that of the pericenter passage, which resulted in many of the stars in the dwarf outskirts to be tidally shocked by the MW.
\end{enumerate}

The first condition is fulfilled if the gas represents $\sim$50\% of the
baryonic mass within the initial half-mass radius of the dwarf
progenitor. For example, such a condition is well reached for the dwarf
irregular WLM \citep{Yang2022a} that has a stellar mass similar to that of Fornax \citep{McConnachie2012}. The second condition is likely reached because there is a significant excess of dwarf spheroidal and ultra-faint dwarfs lying near pericenter \citep{Fritz2018,Hammer2020,Li2021}, where both ram pressure and tidal forces are maximal. \\

Simulations by \citet{Yang2014} have shown that their initial dwarf 3 is able to reproduce the flat velocity dispersion radial profiles of UMi or Draco. It also reproduces morphologies and surface-brightness of the two dwarfs (see also Figs. 10 and 11 of \citealt{Hammer2018a}). In these simulations a significant part of the initial stars has been lost during the dwarf infall in the MW halo, due to their expansion after gas removal.\\

 \begin{figure}
\includegraphics[width=3.5in]{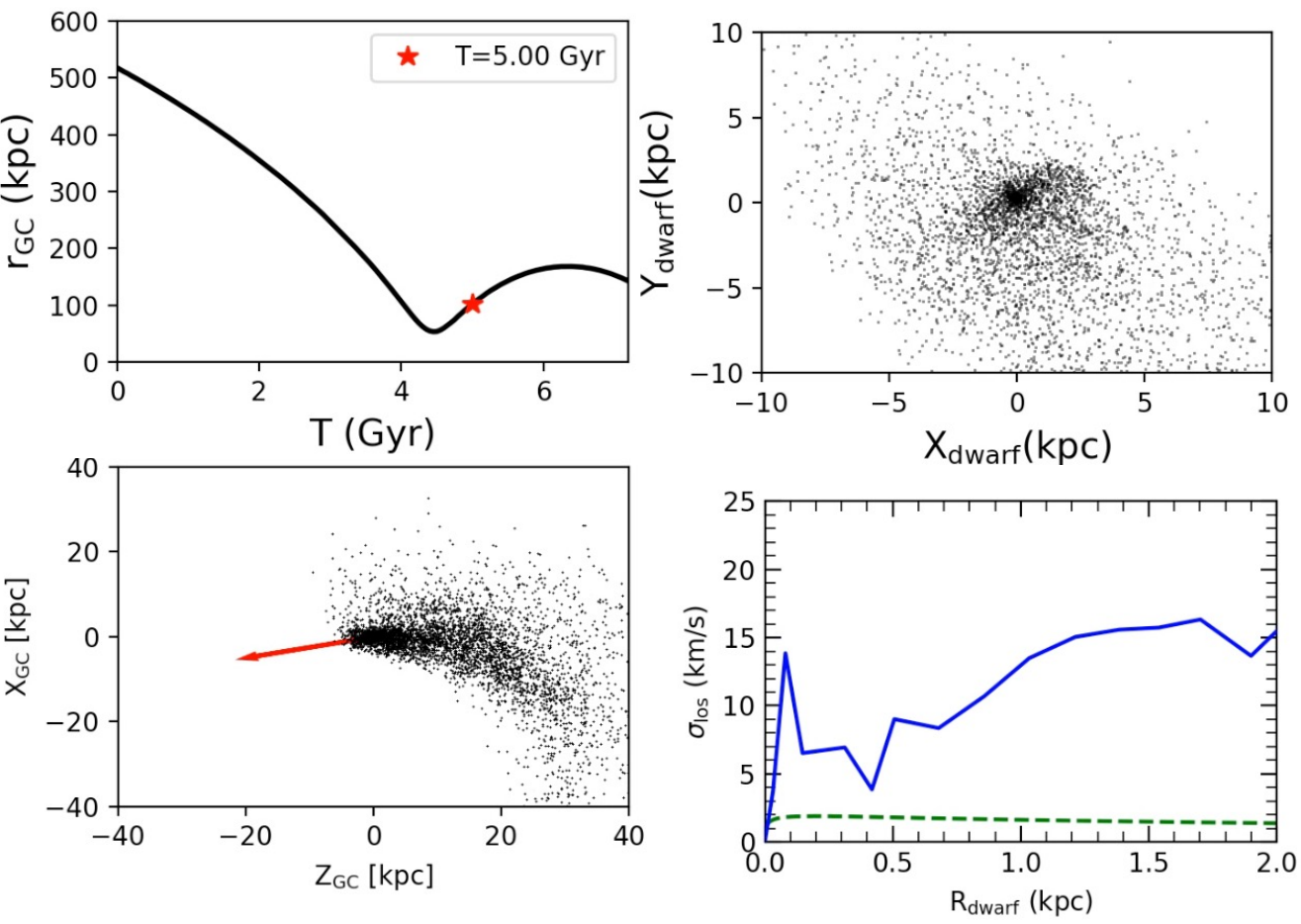}
\caption{Evolution of stars in a hydrodynamical
  simulation made in Paper~III and based on a MW mass model of $8\times 10^{11}$ $\msun$, with a
  halo gas mass density following that has been adopted by \citet{Wang2019}
  to reproduce the Magellanic stream. In this simulation the dark matter halo
  is truncated at $r_{200} = 189\,\rm kpc$. \emph{Top-left panel} gives the trajectory of the dwarf (Galactocentric distance versus time) that
   follows that of Sculptor, whose star properties are captured 0.5 Gyr after pericenter passage  (\emph{see the red point in the top-left panel}). \emph{Top-right} panel indicates that a significant fraction of stars have left the main body of the dwarf while \emph{bottom-left panel} shows that it brings an additional extended component mostly along the line of sight (indicated by the \emph{red
    arrow}). \emph{Bottom-right panel} shows the radial profile of the line-of-sight velocity-dispersion similarly to measurements currently made for MW dwarfs (\emph{blue solid line}) and compare it to expectations from equilibrium with its stellar mass content (\emph{green dash line}). More details will be given in Paper~III. } 
\label{fig:simulations}
\end{figure}

 More simulations are needed to verify whether each MW dwarf can be
 reproduced in this way, including the more massive ones, e.g., Sculptor (see Paper~III), as well as ultra-faint dwarfs. One difficulty is numerical, e.g., it is essential to ensure a high resolution and thus small mass for the stellar and gas particles in the dwarfs, the latter of which have to interact with MW hot gas particles, the total mass of which is a hundred to a thousand times larger than that of the initial dwarfs. Comparison between simulations and observations needs to consider all stellar particles, including those that are just in projection on the dwarf core since they also contribute to $\sigma_{\rm los}$ (see, e.g., bottom-left panel of Figure~\ref{fig:simulations}).\\
 
Simulations of Paper III show that if the above conditions (i), and (ii) are granted, the velocity dispersion (bottom-right panel of Figure~\ref{fig:simulations}) is much higher than expectations from self-equilibrium of the stellar component alone (compare the dashed green line with the solid blue line). This increase is found maximal at pericenter passage (T$\sim$ 4.5 Gyr, see top-left panel of Figure~\ref{fig:simulations}), while such a phenomenon has a duration of several hundred of million years. This is indicated by Figure~\ref{fig:simulations} that represents the stellar particle properties at T= 5 Gyr (see the red star in the top-left panel), when velocity dispersion is still considerably increased (compare blue solid and green dash lines in bottom-right panel). This additional velocity dispersion is provided by stars unbound to the dwarf, which are affected by tidal shocks as expected from Eq.~\ref{fig:ffs_gMW}. The situation in the central core of the dwarf is probably more complex, since $\sim$300 Myr after the gas release, there are still stars that are in expansion because of their large velocity, while some other core stars can also be affected by tidal shocks since their small velocity may place them in the impulse approximation conditions.\\
  
\subsection{Combination of tidal shocks and ram-pressure stripping}
\label{sec:comb}
Figure~\ref{fig:simulations} also illustrates that Equation~\ref{Eq:ffs_gMW} is too simplistic in assuming that all the  energy exchange ($\Delta E = \sigma^2$/2) occurs only along the line of sight, especially if the main contribution occurs at pericenter. In reality, the exchange of energy should result from an integration over all the past orbit of the dwarf galaxy. It would affect the validity of Equation~\ref{Eq:ffs_gMW}  especially for nearby dwarf galaxies, for which the direction of $g_{\rm MW}$ when calculated at pericenter may differ from the line-of-sight. This could explain why the correlation slope in Figure~\ref{fig:ffs_gMW} is smaller than 1, i.e., because the X-axis value could have been overestimated for nearby dwarfs, such as Segue I or Grus I that lie on the right of the Figure.  \\

Another limitation of  Equation~\ref{Eq:ffs_gMW} 
is that in addition to MW tidal shocks, gas removal also leads to an
increase of the dwarf velocity dispersions. This is because after gas loss,
stars keep the memory of their initial velocity
dispersion, which is large because it balanced the initial total mass of both
gas and stellar components. Since ram pressure stripping is also expected to be maximal near pericenter, it could contribute to the correlations found in Figures~\ref{fig:ffs_Rper} and ~\ref{fig:ffs_gMW}, in addition to tidal shocks described by Equation~\ref{Eq:ffs_gMW}.\\
Paper~III shows that this could explain the
larger velocity dispersion of Antlia II when compared to that of Crater
II. The latter, having not passed its pericenter, is mostly affected by the
gas removal, while the former, having passed its pericenter, is
furthermore affected by tidal shocks. Reproducing the excess of velocity
dispersion in all dwarfs, including ultra-faint dwarfs, would require a
specific modeling of each of them. Besides numerical limitations discussed
above, this is also complicated by the fact that new observations from deeper
surveys (e.g., from \citealt{Cantu2021} and \citealt{Chiti2022}) have shown
that structural parameters such as the half-light radius of Grus I may have
changed from 28 to 151 pc. On the other hand, ultra-faint dwarf orbits
often have larger eccentricity than classical
dwarfs, which suggests a better efficiency of MW tidal shocks. 
 
\subsection{Consistency of dwarf infall times with their star formation histories}
\label{sec:disc_SFH}
A scenario for which MW halo gas and gravity has recently shaped the dwarf morphologies and kinematics differs with results of many previous studies, for which constraints about infall times of MW dwarf galaxies were mostly coming from their star formation histories.  \\

While some classical dwarfs have extended star-formation histories (Fornax,
Carina, LeoI, Leo II, Canes Venaciti I), some other (Sculptor, Sextans, Ursa Minor,
and Draco) show only very old stellar populations \citep[see, e.g.,][]{Weisz2014}. Many ultra-faint dwarfs share the
latter property, though the result is less robust given the
lack of RGB stars to determine age and metal abundances (Vanessa~Hill, 2019,
private communication). This has led some studies to assume very early infall
events for most dwarfs, even reaching the ionization epochs \citep[and
  references therein]{Seo2023}. It has also been attempted to reconcile
infall times with star formation histories
\citep{Rocha2012,Fillingham2019,Miyoshi2020,Barmentloo2023}, on the basis
that the dwarf gas is likely to be stripped during the infall, or
alternatively, could be removed by other mechanisms (e.g., 
feedback from supernovae and mergers) at very early epochs. \\

However, star formation histories may not unequivocally trace the orbital history. As a first counter example,  Draco, Ursa Minor, Carina, and Canes Venaciti I share similar stellar mass and orbital energy within 0.2 dex (factor 1.6), which suggests similar infall times. However, the former two show no star formation since almost 10 Gyr, and the two latter have a star formation still active 1.5-2 Gyr ago \citep{Weisz2014,Martin2008}. A second example is provided by ultra faint dwarfs that may have not been able to form stars before their infall to the MW halo because of their too small gas surface density, which is likely below the Schmidt-Kennicutt law \citep{Kennicutt1998}. \\

Predicting the orbital history from the star formation history requires accounting for the very last star formation event, even if it corresponds to a very small fraction of the stellar mass. Fornax provides a good illustration of this, because \citet{deBoer2013} found a new stellar over-density, located 0.7 kpc from the centre, which is only 100 Myr old, but accounts for a tiny fraction of the stars. This indicates that the last part of its gas has left Fornax very recently\footnote{Part of this gas may have been directly detected as a very large HI gas cloud superposed on Fornax \citep{Bouchard2006}.}, likely through ram-pressure exerted by the MW halo hot gas. 
A recent gas removal is consistent with a recent first infall for which the ram-pressure may have slowed it down reducing its orbit eccentricities (see Paper~III). Conversely, a first infall 8 Gyr ago is unlikely, because Fornax would have accomplished about 4 pericenter passages since then which should have removed the gas much earlier than 100 Myr ago\footnote{This is because the gas of the MW has been already tested at large distance from the modeling of the Magellanic Stream, from which it should reach density of $10^{-4}$ atoms per $cm^3$ at 100-200 kpc \citep{Hammer2015,Wang2019}, values sufficiently high to strip Fornax gas in less than 2 orbits.}.\\

Having many dwarfs entering the MW halo 8-10 Gyr ago is in sharp
contradiction with the orbital energy-infall time correlation shown in fig.~1
of \citet[see also fig.~6 of Paper~I, as well as Appendix~\ref{sec:Barmentloo} and Figure~\ref {fig:Barmentloo}]{Rocha2012}. This is because such an epoch coincides with that of the GSE event, which shows an orbital energy 5 (0.7 dex) times smaller than the average energy of dwarfs (see Appendix~\ref{sec:Barmentloo}). \\

A recent infall of most MW dwarf galaxies predicts:
\begin{enumerate} 
\item Less than 3 Gyr ago they were gas-rich and they have lost their gas near their first pericenter passage;
\item Before being removed, the gas is pressurized by the MW hot corona, which leads to star formation, and at least a small fraction of young stars is expected to be found in their cores;
\item The fraction of young stars depends on whether young stars are kept in the central core or are expanding in the dwarf outskirts.
\end{enumerate} 
For Fornax, \citet{deBoer2012} measured that only a few percent of stars are
younger than 2 Gyr. The most interesting dwarfs to test are Sculptor, Ursa
Minor, and Draco, for which color-magnitude diagrams (CMDs) are sufficiently
populated \citep{deBoer2011,Munoz2018}, and that are fully dominated by old
stars \citep{deBoer2012,Weisz2014}.  Yang et al. (2023, in preparation,
hereafter Paper~IV) has re-analyzed CMDs of these classical dwarfs among
others. They found evidence for massive and young stars that lie on the blue
side of the RGB branch. In particular, Sculptor,
Ursa Minor, and Draco contain a small fraction of young stars, implying
recent gas loss, consistent with a recent infall less than 3 Gyr
ago. Simulations of Paper~III predict that young stars formed during a
ram-pressure event have their motions strongly affected by the gas that is
leaving the dwarf. Consequently, these simulations of Paper~III suggest that only a small fraction of young stars stay in the core, in agreement with observations (see Paper~IV).  

\subsection{Predictions of stellar halos surrounding dwarf galaxies}
\label{sec:disc_outskirts}

The scenario of a recent accretion of gas-rich MW dwarfs, accompanied by a
recent gas removal, followed by stellar expansion, and efficient MW tidal
shocks, predicts that dwarf outskirts should be populated well beyond their
half-light radii.
Simulations of such a combination of effects show (see
  Fig.~\ref{fig:simulations}) that a significant part\footnote{ Even without recent gas stripping and tidal shocks, there is always a floor
      non-zero value of stars evaporating in a self-gravitating system. In the future on may compare predictions of Figure~\ref{fig:simulations} with simulations of 
      systems lacking gas (no gas removal effect) and in circular orbits of radius equal 
      to the present galactocentric distances (no tidal shocks).} of the initial stellar
content of the dwarf galaxy is expelled into the MW halo (see
section~\ref{sec:escapingstars}). Figure~\ref{fig:simulations} predicts that
most dwarfs affected by this effect should be accompanied by expanding stars
in their outskirts, while the distance they have covered depends on the epoch
when the gas has been removed and if the outer dark matter could have been
  tidally stripped after a single passage. \\

While investigations of dwarf outskirts is a very novel field mostly based on
Gaia observations, recent studies have found member stars lying far or very
far from dwarf galaxy cores. Fornax has been intensively studied by
\citet{Yang2022b}, and they found an additional component in its
outskirts. More recently, the combination of Gaia and high resolution
spectroscopy has convincingly detected the presence of stars from 4 to 12
$r_{\rm half}$ in Ursa Minor \citep{Sestito2023}, Ursa Major I, Coma
Berenices, Bootes I \citep{Waller2023}, Tucana II \citep{Chiti2023}, and Sextans \citep{Roederer2023}. In
addition, when observing Grus I by using a much deeper photometry than
\citet{Munoz2018}, \citet{Cantu2021} found a considerably much larger size,
with a half-light radius increased by a factor over five:
passing from 28 pc to 151 pc. \\

Future observations will show whether there are evaporating stars in most MW dwarfs, which is expected if they are expanding stellar systems after gas and gravity losses.

\subsection{Could the correlations be generated by selection effects?}
\begin{figure}
\includegraphics[width=3.5in]{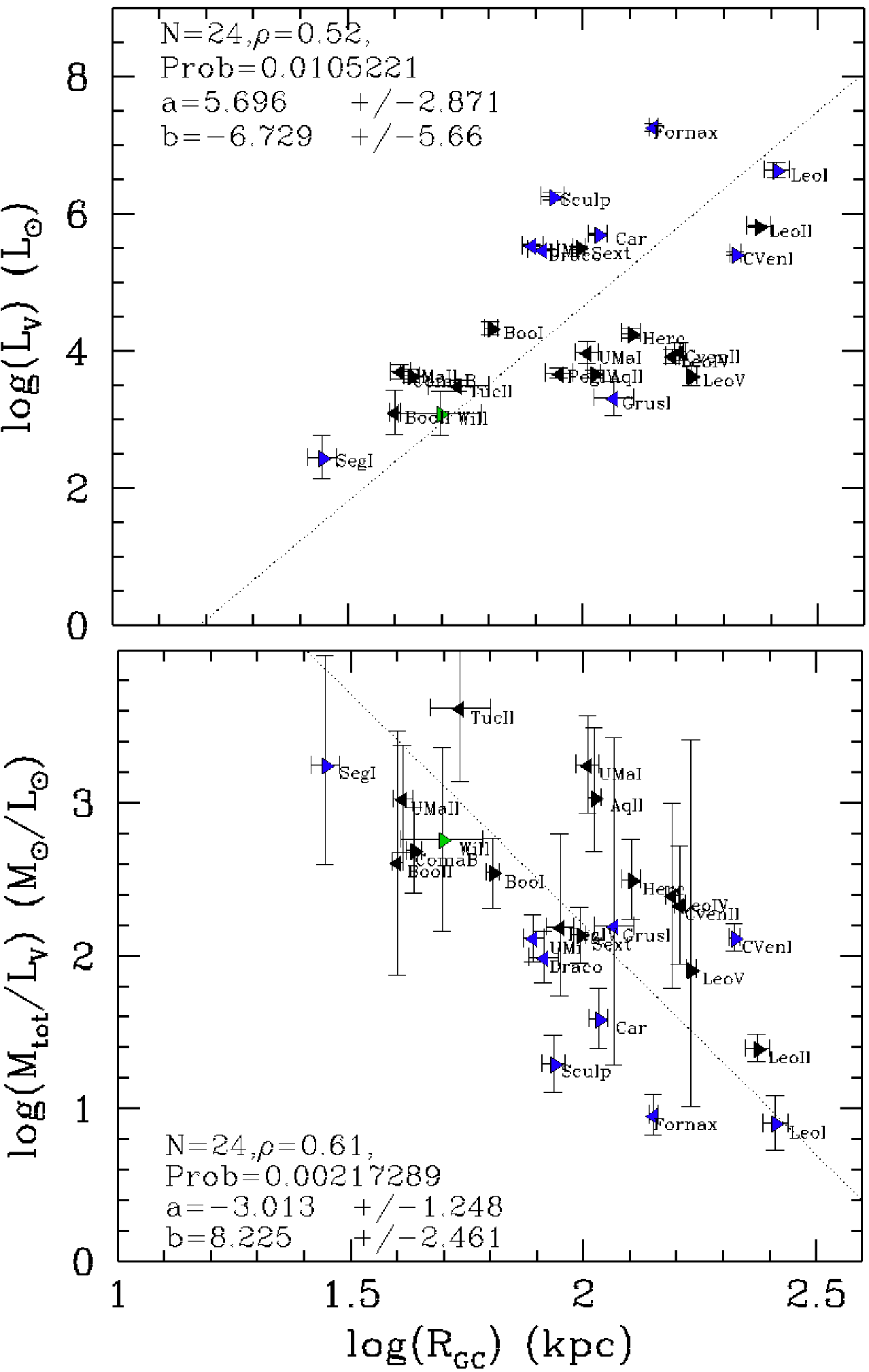}
\caption{Correlation between $V$-band luminosity $L_{V}$ (\emph{top panel})
  and $M_{\rm tot}$/$L_{V}$ (\emph{bottom panel}) with the galactocentric
  radius $R_{\rm GC}$.
 } 
\label{fig:Lv_MsLv}
\end{figure}

The strong correlations found in Figures~\ref{fig:ffs_Rper} and ~\ref{fig:ffs_gMW} are likely responsible for most correlations shown by dwarf galaxies all along this paper. Their impact is strong because they correspond to a theoretical prediction (see Eq.~\ref{Eq:ffs_gMW}), for which MW dwarf galaxies are stellar systems out of equilibrium due to the severe impact caused by the gas removal and MW tidal shocks. However, one may wonder if there could be selection effects, e.g., due to the fact that not all the MW dwarfs have been discovered, or for which their velocity dispersion has not been determined, yet. \\

The top panel of Figure~\ref{fig:Lv_MsLv} reveals the trend between dwarf visible luminosity and galacto-centric distance. This could be due to selection effects as faint dwarfs are more difficult to be detected at larger distances \citep{Drlica-Wagner2020}. 
 This may generate the anti-correlation between $M_{\rm tot}/L_{\rm V}$ and
 $R_{\rm GC}$, because $M_{\rm tot}$ has been calculated from Eq.~\ref{Wolf},
 and does not correlate with $R_{\rm GC}$ (see red-dashed lines in
 Figure~\ref{fig:sig2stars_rhX_Rp} for $X=1$).\\
 
\begin{figure}
\includegraphics[width=3.5in]{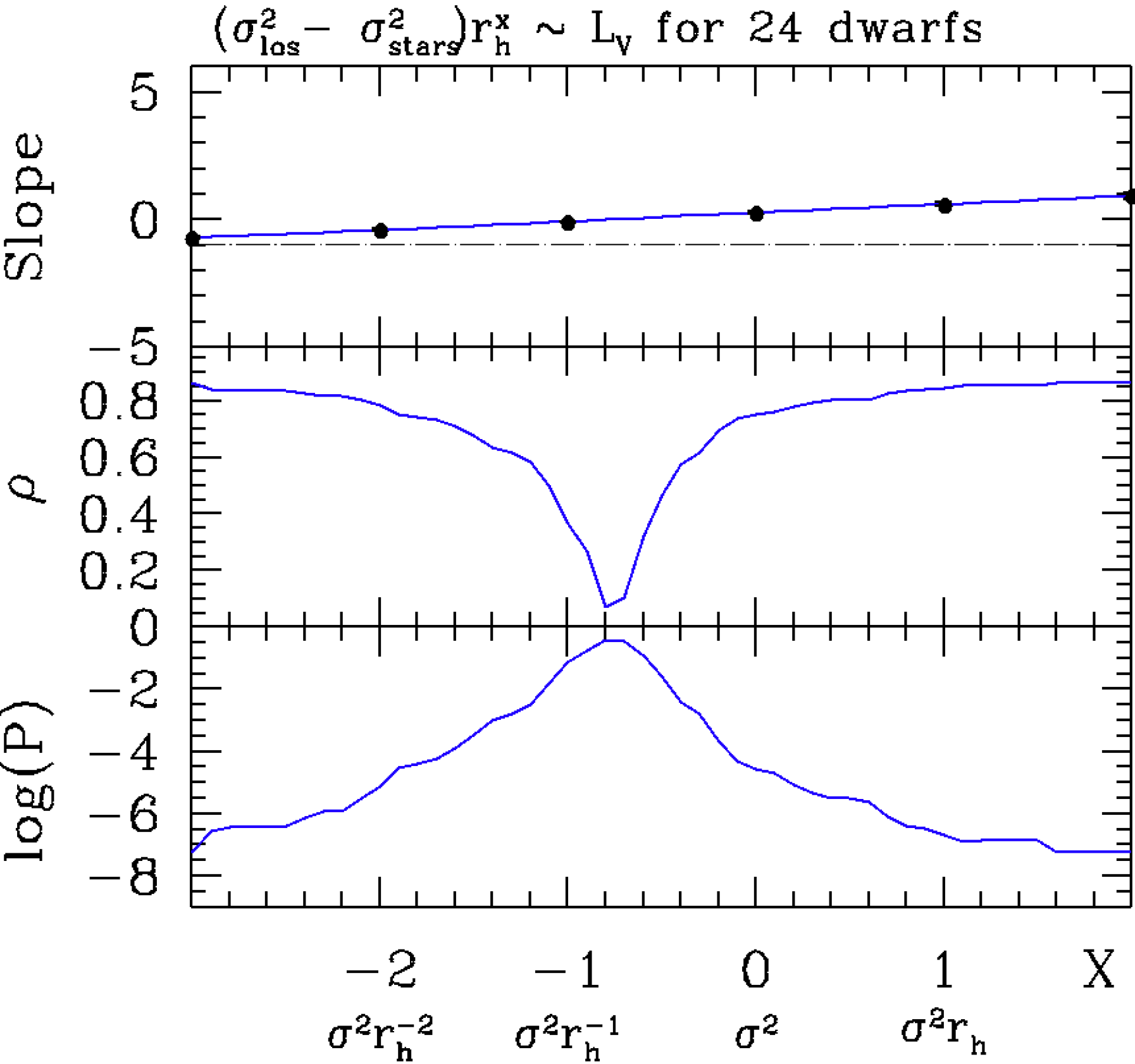}
\caption{Slope, correlation $\rho$, and associated probability (in
  logarithmic scale) of the correlation between $\sigma_{\rm los}^2 r_{\rm
    half}^X$ and $L_{V}$  from $X=-3$ to $X=2$. Each quantity is shown on the
  $X$ axis.
}  
\label{fig:sig2stars_rhX_Lv}
\end{figure}

One may wonder whether or not this possible selection effect can also be responsible of the correlations found in this paper. Figure~\ref{fig:sig2stars_rhX_Lv} provides a negative answer, because the structural parameter that is the most correlated with the pericenter, $(\sigma_{\rm los}^2 - \sigma_{\rm stars}^2)$/$r_{\rm half}$, is not correlated with $L_{V}$. It results that the correlations between  $(\sigma_{\rm los}^2 - \sigma_{\rm stars}^2)$/$r_{\rm half}$ and $R_{\rm peri}$ or $g_{\rm MW}$ are unlikely to be affected by selection effects.

\subsection{Are MW dwarf dark-matter contents over-estimated?}
\label{sec:disc_DM}

A recent arrival of most dwarfs in the MW halo, together with recent bursts of star formation, favors a scenario in which they have recently lost their gas, making them unstable against MW tides. The fact that most dwarfs are near their pericenter also supports the recent loss of gas together with MW tidal shocks. This is indicated by the correlations shown in Figures~\ref{fig:ffs_Rper} and ~\ref{fig:ffs_gMW}, which are theoretically predicted, as well as reproduced by simulations of Paper~III.\\

However, we notice that the recent infall of MW dwarfs is justified by orbital arguments, with a timing scale coming from comparisons to former merger events in the MW halo (e.g., GSE and Sgr). This would have considerable consequences on the DM content of dwarf progenitors and then on that of MW dwarfs. As shown by \citet{Mayer2006}, it would require several pericenter passages and a Hubble time to transform a DM-dominated rotating gas-rich dwarf into a gas-free, dispersion supported system. Here, this transformation needs to be done in just one orbit, which considerably limits the possible amount of DM. Paper~III has simulated a DM-dominated progenitor of Sculptor, assuming the prescription from Eq.~\ref{Wolf}, which has been established by \citet{Wolf2010}. It confirms that in the Sculptor orbital conditions, a rotating gas-rich dwarf cannot be transformed into a MW gas-free dwarf after one orbital time.

This scenario implies that most MW dwarfs as they are observed today are not in equilibrium. Their recent gas removal together with MW tidal shocks suffice to explain their high velocity dispersions, both from theory and simulations. It results that the self-equilibrium conditions assumed by \citet{Walker2009} and \citet{Wolf2010} cannot apply to MW dwarf galaxies, which questions the corresponding estimates of large mass excess when they are compared to the stellar mass. Consequently, this means that we have no way to prove or disprove the presence of dark matter in objects far from equilibrium. 







\section{Conclusions}
\label{sec:conclusions}

{Here, we have investigated how the
    structural (morphological and kinematical) properties of Milky Way globular clusters and dwarf
    spheroidals depend on their orbital properties. We have further limited our investigations to relations that do not depend on the adopted MW mass.
We first confirm that the $r_{\rm half} \propto 1/E$ relation found in
    Paper~I is robust to the adopted Milky Way mass model.\\
  
We also confirm that HSB-GCs are in pseudo-equilibrium with MW tidal
shocks, removing approximately 3\% of their mass at each pericenter
passage. They differ from the more fragile LSB-GCs, which are strongly
destabilized by MW tides, a significant fraction of them (27 \%) possess
tidal tails, and/or are in a tidal dominant regime (24\%, see
Figure~\ref{fig:tidalshocks}).

Dwarf galaxies show some similarities with LSB-GCs. However, correlations
between their structural and orbital properties are unexpected if they
arrived recently into the MW halo and have no time to perform more than one
orbit (see Paper~I). Specifically the anti-correlation shown in
Figure~\ref{fig:ffs_Rper} or that found by \citet[see their
  Fig. 1]{Kaplinghat2019} is in tension with MW dwarfs modeled as systems at
self-equilibrium with large amounts of dark matter}. Indeed, it is difficult to explain why new-coming sub-halos have their densities depending on pericenter without having time to be affected by MW tides \citep{Cardona-Barrero2023}. \\



A late infall for dwarfs requires one to consider the properties of
their pre-infall progenitors outside the MW halo. They are likely gas-rich
dwarf galaxies \citep{Grcevich2009}, and their passage into the MW halo gas
may have fully transformed them into gas-free dwarf galaxies
\citep{Mayer2006,Yang2014}. It suggests that most dwarf properties result
from a temporal sequence, beginning with gas stripping due to MW halo gas
ram-pressure, expansion of their stars due to the subsequent lack of gravity,
and then a significant impact of MW tidal shocks exerted mostly on the
leaving stars. The impact of such an out-of-equilibrium process has been
theoretically described, and will be shown in Paper~III simulations, for which a first example is provided in Figure~\ref{fig:simulations}. It is also consistent
with the dwarf proximity to their pericenters, a property that otherwise
would appear in contradiction with conservation of
  energy. \\


This scenario allows us to make the following predictions: Most MW dwarf galaxies 
 (1) have velocity dispersion values with a significant contribution due to ram-pressure stripping and Galactic tidal shocks; (2) show many stars in their outskirts, their distances from the dwarf core depending on the elapsed time since the gas has been decoupled from stars; (3) show a small fraction of young stars, both in their cores and outskirts.  \\
 
Condition (1) is fulfilled through Eq.~\ref{Eq:ffs_gMW} and Figure~\ref{fig:ffs_gMW}, as well as from simulations. Verification of condition (2) is on-going through the recent and successful discoveries of stars in the very outskirts of dwarfs \citep{Sestito2023,Waller2023,Chiti2023,Cantu2021,Yang2022b,Roederer2023}. Paper~IV reveals that condition (3) is also fulfilled on the basis of a novel investigation of their CMDs. This series of papers (Papers~I to IV) may change our understanding of MW dwarf galaxies, passing from an equilibrium model allowing mass estimates, to an out-of-equilibrium model that prevents mass estimates other than that of the baryonic mass.\\

Further simulations could be useful to verify which amount of dark matter they may contain for allowing their transformation into gas-free, dispersion-supported dwarfs in a single orbit, while preserving the correlations shown in Figures~\ref{fig:ffs_Rper} and ~\ref{fig:ffs_gMW}.


\section*{Acknowledgments}

  We warmly thank the referee, Dr Luis Alberto Aguilar, for his very useful report, from which we confess having adopted some sentences in the final version, because they better convey the physics underlying this paper.
 We are grateful for the support of the International Research Program Tianguan, which is an agreement between the CNRS in France, NAOC, IHEP, and the Yunnan Univ. in China. J.-L.W. acknowledges financial support from the China Scholarship Council (CSC) No.202210740004, as well as Y.-J.J. (No.202108070090).
 Marcel S. Pawlowski acknowledges funding of a Leibniz-Junior Research Group
 (project number J94/2020).

\section*{Data Availability}

All necessary data used in this paper are available in the Tables of the Appendix~\ref{sec:parameters} and in \citet{Li2021}.



\bibliographystyle{mnras}
\bibliography{references_wAbbrevs} 




\appendix
\section{About the infall time of dwarf galaxies}
\label{sec:Barmentloo}

\begin{figure}
\includegraphics[width=3.5in]{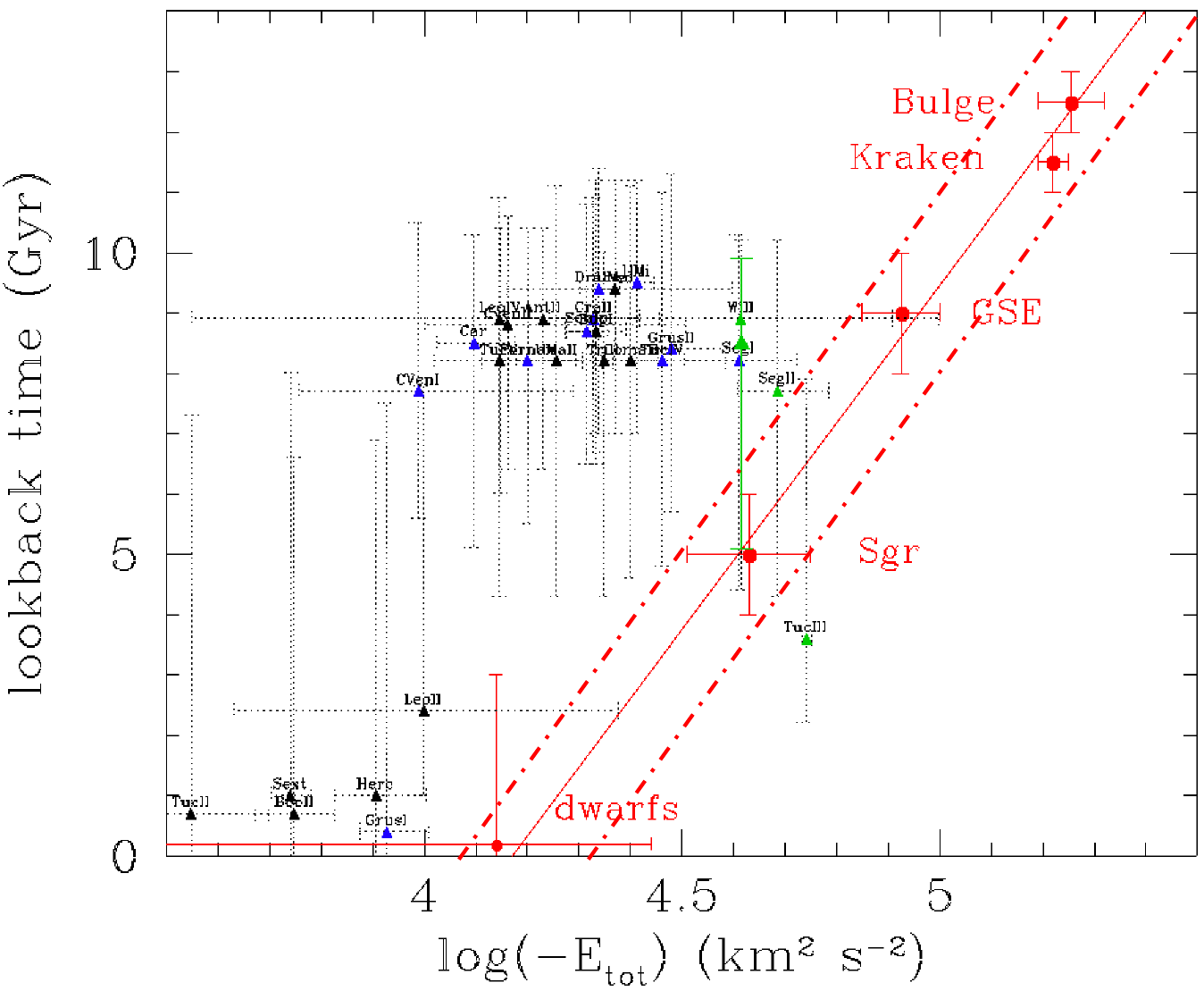}
\caption{Comparison of the infall lookback time as a
  function of the orbital energy from
  \citealt[\emph{red dots} and \emph{line} joining MW merger
    events]{Hammer2023}, which matches very well \citealt[the \emph{two red dot-dashed
      lines} representing $\pm$$1\,\sigma$]{Rocha2012}. Besides this analysis based on a strong correlation between satellite orbital energy and infall time, \emph{triangles} represent individual dwarfs infall times based on the method of \citet[see text]{Barmentloo2023}, showing how these predictions are discrepant from the expected relation with total orbital energy.  The dwarfs associated with the VPOS and with Sgr
    are shown in \emph{blue} and \emph{green}, respectively. 
} 
\label{fig:Barmentloo}
\end{figure}

\citet{Barmentloo2023} proposed a machine learning technique to calculate the
infall time of most MW dwarfs. Figure~\ref{fig:Barmentloo} compares their
results for 25 dwarfs. This Appendix examines the robustness of their
analysis, and compares it to results from cosmological simulations
\citep{Rocha2012}. We make the following points.
\begin{itemize}
\item \citeauthor{Barmentloo2023} find an infall epoch for Sgr (and other associated dwarfs, see green points) that is later than that of VPOS dwarfs, while the Sgr system has a much smaller energy (see Figure~\ref{fig:Barmentloo}). This contradicts their own claim, i.e., quoting them, "we expect that there is a strong correlation between satellite orbital energy and infall time." 
\item The infall time for most dwarfs would be similar to that of the GSE major merger event in the MW, while the orbital energy of the later event is five times smaller;
\item The use of non-independent input features (e.g., distance, radial and total velocity versus total energy and angular momentum) to derive the infall time may lead to unreliable results, because some of them show a flat profile versus the infall time, i.e., which likely dilutes the predictions;
\item According to their fig.~4, their technique can only retrieve input infall times smaller than 3-4 Gyr, i.e., only for dwarfs shown in the bottom-left part of Figure~\ref{fig:Barmentloo};
\item Their choice of host galaxy halos in the EAGLE simulation appears not representative of the MW past history, i.e., the significant increase of both pericenter and apocenter from 9 to 1 Gyr ago (see their Fig. 1) requires considerable mass gains, while MW is known to have experienced its last major merger 9-10 Gyr ago, as shown from the analysis of the GSE event \citep{Naidu2021};
\end{itemize}

Additionally, the choice of \citet{Barmentloo2023} to define the infall time as being the first time the future satellite is passing through the virial radius of the host lead to very scattered results according to \citet{Rocha2012}. Consequently, predictions for infall times are better predicted through their tight correlation with orbital energy \citep{Rocha2012}, the latter being calculated with good accuracy from Gaia DR3 (see Paper~I). \\

We also noticed the study by \citet{Pagnini2023}, which may cast some doubts
about the reliability of associating GCs to past merger events in the MW
\citep{Kruijssen2019,Kruijssen2020,Malhan2022}. However, this might be due to
the following assumptions (1) that GCs originate in halos of infalling dwarfs
and are not formed during merger events (e.g., Kraken GSE, or Pontus) that
likely induced strong star formation events 12 to 9 Gyr ago
\citep{Haywood2016}, (2) that all mergers experienced by the MW, including
GSE, were minor (e.g., mass ratio of 1:10), contrary to the analysis by \citet{Naidu2021} who considered mass ratio from 1:2 to 1:4\footnote{Notice that these higher mass ratios are necessary to explain the origin of both thin and thick disk of a spiral galaxy like the MW \citep{Hammer2009,Hammer2018b,Hopkins2010, Sauvaget2018}.} and (3) that simulations without gas can reproduce the infall of stellar systems at epochs when the gas is preponderant. It is unlikely that all GCs have been formed through the way proposed by  \citet{Pagnini2023}. However, the latter study provides a complementary channel for explaining several GCs that are not identified inside a structure in the plane made by total energy and angular momentum (see Fig. 5 of Paper~I).

\section{Intrinsic parameters of MW dwarf galaxies}
\label{sec:parameters}

Table~\ref{tab:struct} describes the structural parameters of the MW dwarf galaxies. Column 1: dwarf galaxy name; Column 2: V- luminosity; Column 3: stellar mass to light ratio; Column 4: Galacto-centric distance; Column 5: half-light radius or effective radius; Column 6: dwarf ellipticity; Column 7: line of sight velocity dispersion; Column 8: velocity dispersion due to the sole stellar component.  \\

Data of Table~\ref{tab:struct} are taken from the review
by \citet[see also references therein]{Simon2019}, and have been updated by more recent measurements. The latter include:
\begin{itemize}
\item New estimates of $\sigma_{\rm los}$ for Bootes I, Leo IV, and LeoV \citep{Jenkins2021};
\item Last update by Josh Simon of \citet{Simon2019} with a new value for $\sigma_{\rm los}$ of Grus I, and Leo V;
\item New measurements from \citet{Bruce2023} of $\sigma_{\rm los}$ for Aquarius II (8 spectroscopic stars, 4.7 instead of 5.4) and Bootes II (2.9 instead of 8.2!); \item First robust measurements of $\sigma_{\rm los}$ of Pegasus IV \citep{Cerny2023};
\item Data for both Antlia II and Crater II \citep{Caldwell2017,Torrealba2016,Torrealba2019,Ji2021}.
\end{itemize}

The sample of dwarf galaxies comes from table~1 of \citet{Li2021} for  46 dwarfs. Here, we only include objects within 300 kpc (excluding Eridanus II), and for which a measurement of $\sigma_{\rm los}$ has been performed without ambiguity. The latter condition leads to remove 5 dwarf galaxies having less than 5 stars with both Gaia and spectroscopy data. It would lead to 40 dwarfs, to which we have further removed the 3 potential GCs (Crater, Draco II, and Sgr II), and 5 dwarfs (Carina II, Carina III, Phoenix II, Horologium I, Hydrus I, and Reticulum II) associated to the LMC.  Also associated to the LMC, Carina III is already excluded since only 4 of its stars possess spectroscopy. Similarly, Columba I (3), Horologium II (1), Pisces II (3), and Reticulum III (3) are not considered due to their lack of spectroscopic stars (which numbers are given in parenthesis). Finally, we have also removed Grus II, Hydra II, Segue 2, Triangulum II, Tucana III, Tucana IV, and Tucana V, because only a limit on their velocity dispersion can be determined \citep[see also references therein]{Simon2019}. 

However, in this paper, we have reintegrated Aquarius II (8 spectroscopic stars) and Pegasus IV, since for both galaxies their velocity dispersion has been measured. It leaves us with a sample of 26 galaxies, all with more than 10 spectroscopic stars, except for Aquarius II, Grus I, Leo IV, Leo V, Pegasus IV, Ursa Major II, and willman.

\onecolumn
\centering
\begin{longtable}{llcccccc}
\label{tab:struct}\\
\caption{Structural properties of Milky Way dwarfs including kinematics.  }\\
\hline
\hline
name & $L_{\rm V}$ & $M_{\rm stars}/L_{\rm V}$ & $R_{\rm GC} $ & $r_{\rm half} $ & $\epsilon$ & $\sigma_{\rm los} $ & $\sigma_{\rm stars}$\\
	 & $(L_{\rm sun}$) &  & (kpc)& (pc) &  & (km s$^{-1}$) & (km s$^{-1}$) \\
\hline
\vspace{-8pt}
\endhead
\hline
\endfoot
Antlia II&$214,783 \pm 29,600$&$2.5$&$130.35^{+6.5}_{-6.5}$&$2867^{+312}_{-312}$&$0.38$&$5.71^{+1.08}_{-1.08}$&$0.317 \pm 0.039$\\
Aquarius II&$4742 \pm 613.2$&$2.5$&$105.40^{+3.3}_{-3.3}$&$160^{+26}_{-26}$&$0.39$&$4.7^{+1.8}_{-1.2}$&$0.2 \pm 0.029$\\
Bootes I&$21,880 \pm 5082$&$2.5$&$63.64^{+2}_{-2}$&$191^{+8}_{-8}$&$0.30$&$5.1^{+0.6}_{-0.7}$&$0.392 \pm 0.054$\\
Bootes II&$1282 \pm 955$&$2.5$&$39.82^{+1}_{-1}$&$39^{+5}_{-5}$&$0.25$&$2.9^{+1.6}_{-1.2}$&$0.210 \pm 0.092$\\
CanesVenatici I&$265,500 \pm 14,680$&$2.5$&$210.8^{+6}_{-6}$&$437^{+18}_{-18}$&$0.44$&$7.6^{+0.4}_{-0.4}$&$0.90 \pm 0.043$\\
CanesVenatici II&$10,000 \pm 2990$&$2.5$&$160.6^{+4}_{-4}$&$71^{+11}_{-11}$&$0.40$&$4.6^{+1}_{-1}$&$0.435 \pm 0.098$\\
Carina&$515,200 \pm 23,740$&$2.5$&$107.6^{+5}_{-5}$&$311^{+15}_{-15}$&$0.36$&$6.6^{+1.2}_{-1.2}$&$1.49 \pm 0.070$\\
ComaBerenices&$4406 \pm 1023$&$2.5$&$43.19^{+1.55}_{-1.55}$&$69^{+4.5}_{-4.5}$&$0.37$&$4.6^{+0.8}_{-0.8}$&$0.29 \pm 0.0435$\\
Crater II&$162,900 \pm 15,030$&$2.5$&$116.4^{+1.1}_{-1.1}$&$1066^{+86}_{-86}$&$0$&$2.7^{+0.3}_{-0.3}$&$0.453 \pm 0.039$\\
Draco&$304,800 \pm 14,040$&$2.5$&$81.98^{+6}_{-6}$&$231^{+17}_{-17}$&$0.29$&$9.1^{+1.2}_{-1.2}$&$1.331 \pm 0.079$\\
Fornax&$18,540,000 \pm 2,397,000$&$1.5$&$141^{+3}_{-3}$&$792^{+18}_{-18}$&$0.29$&$11.7^{+0.9}_{-0.9}$&$4.34 \pm 0.330$\\
Grus I&$3732 \pm 1192$&$2.5$&$116.2^{+11.5}_{-11.5}$&$28^{+23}_{-23}$&$0.44$&$2.9^{+2.1}_{-1}$&$0.423 \pm 0.241$\\
Hercules&$18,370 \pm 28867$&$2.5$&$126.4^{+6}_{-6}$&$216^{+20}_{-20}$&$0.69$&$5.1^{+0.9}_{-0.9}$&$0.338 \pm 0.042$\\
Leo I&$4,406,000 \pm 1,149,000$&$1.5$&$257.8^{+15.5}_{-15.5}$&$270^{+16.5}_{-16.5}$&$0.30$&$9.2^{+0.4}_{-0.4}$&$3.62 \pm 0.58$\\
Leo II&$673,000 \pm 24,800$&$2.5$&$235.5^{+14}_{-14}$&$171^{+10}_{-10}$&$0.07$&$7.4^{+0.4}_{-0.4}$&$2.3 \pm 0.10$\\
Leo IV&$8472^ \pm 2048$&$2.5$&$154.6^{+5}_{-5}$&$114^{+13}_{-13}$&$0.17$&$3.3^{+1.7}_{-1.7}$&$0.316 \pm 0.056$\\
Leo V&$4446 \pm 1501$&$2.5$&$169.8^{+4}_{-4}$&$49^{+16}_{-16}$&$0.43$&$2.3^{+3.2}_{-1.6}$&$0.349 \pm 0.115$\\
Pegasus IV&$4800 \pm 800$&$2.5$&$89^{+6}_{-6}$&$41^{+8}_{-8}$&$0.20$&$3.3^{+1.7}_{-1.1}$&$0.396 \pm 0.071$\\
Sculptor&$1,820,000 \pm 235,300$&$2.5$&$86.09^{+5}_{-5}$&$279^{+16}_{-16}$&$0.33$&$9.2^{+1.1}_{-1.1}$&$2.96 \pm 0.27$\\
Segue 1&$283 \pm 205$&$2.5$&$27.84^{+2}_{-2}$&$24^{+4}_{-4}$&$0.33$&$3.7^{+1.4}_{-1.1}$&$0.1259 \pm 0.056$\\
Sextans&$322,100 \pm 17,810$&$2.5$&$97.97^{+3}_{-3}$&$456^{+15}_{-15}$&$0.30$&$7.9^{+1.3}_{-1.3}$&$0.97 \pm 0.042$\\
Tucana II&$3105 \pm 575$&$2.5$&$54.24^{+8}_{-8}$&$121^{+35}_{-35}$&$0.39$&$8.6^{+4.4}_{-2.7}$&$0.185 \pm 0.044$\\
UrsaMajor I&$9638 \pm 3443$&$2.5$&$101.9^{+5.85}_{-5.85}$&$295^{+28}_{-28}$&$0.59$&$7^{+1}_{-1}$&$0.2095 \pm 0.047$\\
UrsaMajor II&$5058 \pm 1223$&$2.5$&$40.81^{+1.95}_{-1.95}$&$139^{+9}_{-9}$&$0.56$&$5.6^{+1.4}_{-1.4}$&$0.221 \pm 0.033$\\
UrsaMinor&$349,900^ \pm 16,120$&$2.5$&$77.89^{+4}_{-4}$&$405^{+21}_{-21}$&$0.55$&$9.5^{+1.2}_{-1.2}$&$1.0776 \pm 0.0527$\\
Willman 1&$1236 \pm 909$&$2.5$&$49.62^{+10}_{-10}$&$33^{+8}_{-8}$&$0.47$&$4^{+0.8}_{-0.8}$&$0.224 \pm 0.109$\\
 \end{longtable}

\begin{table}
 \caption{Dwarf eccentricities for the four Milky Way dark-matter mass models of Table~\ref{tab:mass}. The seven dwarfs without estimates of their internal velocity dispersions have their names in italics.}
\label{tab:ecc}
 \begin{tabular}{lcccc}
\hline
  Dwarf & Einasto$_{\rm HM}$ & NFW & Einasto$_{\rm MM}$ &  Einasto$_{\rm LM}$\\
\hline
AntliaII & 0.454 & 0.414 & 0.494 &  1.08\\
 AquariusII & 0.312 & 0.581 &  1.31 & 3.001\\
BootesI & 0.333 & 0.444 & 0.649 &  1.55\\
BootesII & 0.641 &  0.91 & 1.803 & 2.968\\
CanesVenaticiI & 0.599 & 0.659 & 0.829 & 1.301\\
CanesVenaticiII & 0.728 & 0.765 & 0.875 & 0.998\\
Carina & 0.076 & 0.303 & 0.871 & 2.347\\
ComaBerenices & 0.323 & 0.471 & 0.656 & 1.549\\
CraterII & 0.603 & 0.584 & 0.598 & 0.764\\
Draco & 0.413 & 0.456 & 0.573 & 1.217\\
Fornax & 0.361 & 0.262 & 0.325 & 1.099\\
GrusI & 0.821 & 0.913 & 1.446 & 1.795\\
{\it GrusII} & 0.478 & 0.538 & 0.594 & 0.884\\
Hercules & 0.608 & 0.763 & 1.231 & 2.196\\
{\it HydraII} & 1.874 & 3.614 & 5.912 & 11.23\\
LeoI &   0.9 & 1.539 & 1.885 & 3.056\\
LeoII & 0.512 & 0.474 & 0.607 & 0.954\\
LeoIV & 0.594 & 0.745 & 0.899 & 1.268\\
LeoV & 0.967 & 2.438 & 4.469 & 8.846\\
PegasusIV &  0.52 & 0.476 & 0.448 & 0.455\\
Sculptor & 0.323 & 0.365 & 0.518 & 1.308\\
Segue1 & 0.475 & 0.525 & 0.572 & 0.821\\
{\it Segue2} & 0.416 & 0.404 & 0.398 & 0.399\\
Sextans & 0.385 & 0.752 & 1.614 & 3.375\\
{\it TriangulumII} & 0.802 & 0.861 & 0.935 & 1.732\\
TucanaII & 0.682 & 0.955 & 1.872 & 2.982\\
{\it TucanaIII} & 0.873 & 0.881 & 0.886 &  0.91\\
{\it TucanaIV} & 0.357 & 0.429 & 0.521 &  1.07\\
{\it TucanaV} & 0.609 & 0.765 & 1.336 & 2.278\\
UrsaMajorI & 0.318 & 0.247 & 0.272 & 0.565\\
UrsaMajorII & 0.476 & 0.649 & 0.927 & 2.056\\
UrsaMinor & 0.372 & 0.372 &  0.43 & 0.857\\
Willman   &   0.249 & 0.247 & 0.265 & 0.332\\\hline
 \end{tabular}
\end{table}

\newpage 
 
 \section{Properties of pericenter velocity}
\label{sec:Vperi}

\begin{figure}
\includegraphics[width=3.6in]{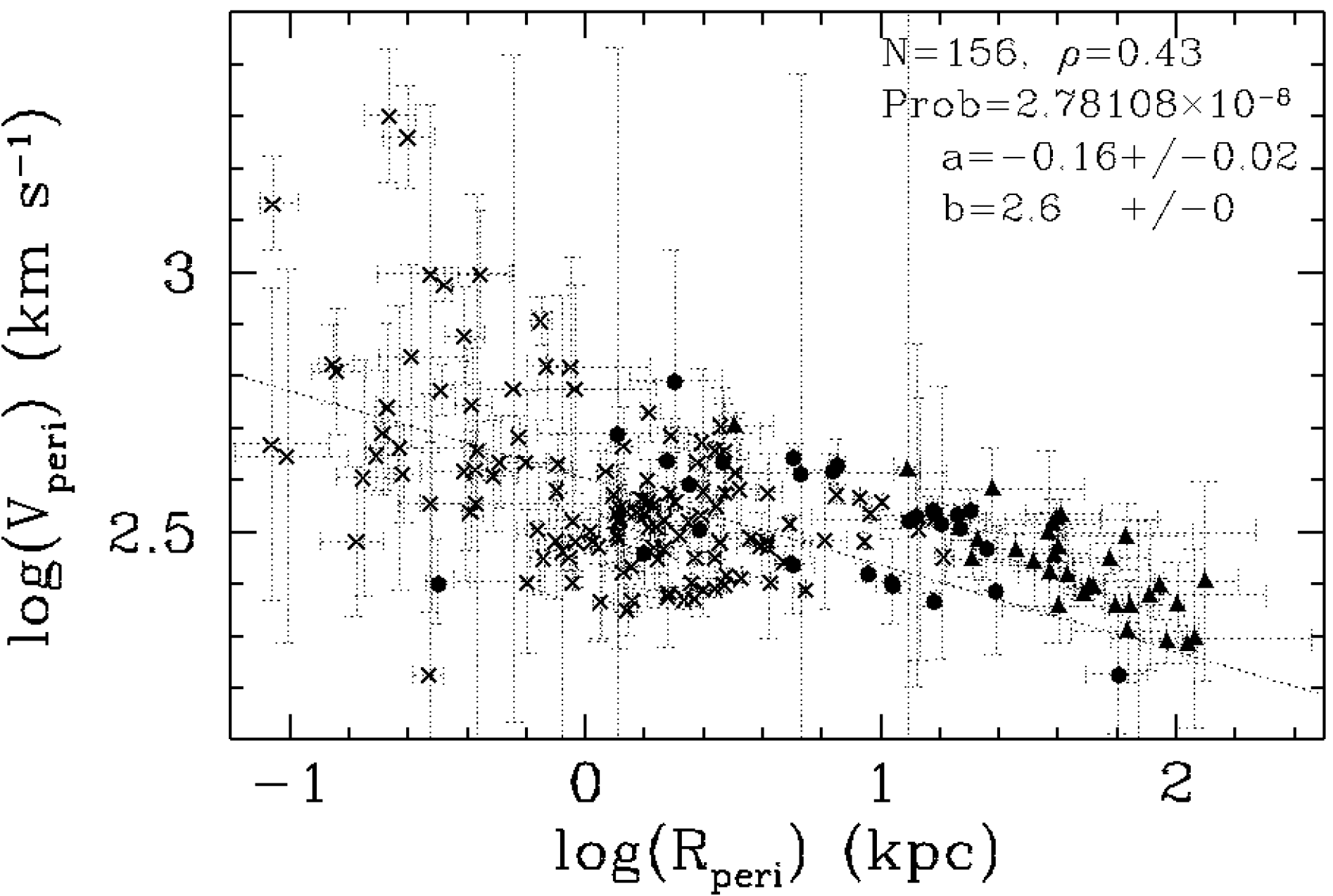}
\caption{$V_{\rm peri}$ versus $R_{\rm peri}$ in logarithmic scale for
  {\rm  HSB-GCs} (\emph{crosses}), {\rm  LSB-GCs} (\emph{full dots}), and dwarfs (\emph{triangles}). The \emph{dotted} line indicates the best fit of the correlation for 156 GCs (see parameters on the top-right).
 } 
\label{fig:Vper_Rper}
\end{figure}

Figure~\ref{fig:Vper_Rper} shows that for GCs, $V_{\rm peri}$ is rather constant (logarithmic slope $-0.16$) with large variations of $R_{\rm peri}$:
\begin{equation}
 V_{\rm peri}= V_{\rm peri} = 398\,\rm km\,s^{-1} \,R_{\rm peri}^{-0.16}
\end{equation}
 The slope is slightly steeper for dwarfs (-0.2), which also show a stronger correlation ($\rho$= 0.9) than GCs.

\bsp	
\label{lastpage}
\end{document}